\documentclass[12pt,a4paper,dvips]{scrartcl}
\usepackage[ansinew,latin1]{inputenc}
\usepackage[T1]{fontenc}
\usepackage{graphicx}
\usepackage{amsmath}
\usepackage{amssymb}

\begin{document}

\begin{titlepage}
\begin{center} {\LARGE \bf Gauge theories as a geometrical issue of a Kaluza-Klein framework
\vspace{0.2in} \\}\vspace*{1cm}
{\bf Francesco Cianfrani*, Andrea Marrocco*, Giovanni Montani*}\\
\vspace*{1cm}
*ICRA---International Center for Relativistic Astrophysics\\
Dipartimento di Fisica (G9),\\
Universit\`a  di Roma, ``La Sapienza",\\
Piazzale Aldo Moro 5, 00185 Rome, Italy.\\
e-mail: montani@icra.it\\
        marroccoandrea@hotmail.com\\
        francesco.cianfrani@icra.it\\
\vspace*{1.8cm}

 \vspace*{1cm} \\

\vspace*{1cm}

{\bf  Abstract\\} \end{center}

We present a geometrical unification theory in a Kaluza-Klein approach that achieve the geometrization of a generic gauge theory bosonic component.\\

We show how it is possible to derive the gauge charge conservation
from the invariance of the model under extra-dimensional
translations and to geometrize gauge connections for spinors, thus we can introduce the matter just by free
spinorial fields. Then, we present the applications to i)a
pentadimensional manifold $V^{4}\otimes S^{1}$, so reproducing the
original Kaluza-Klein theory, unless some extensions related to the
rule of the scalar field contained in the metric and the
introduction of matter by spinors with a phase dependence
from the fifth coordinate, ii)a seven-dimensional manifold
$V^{4}\otimes S^{1}\otimes S^{2}$, in which we geometrize the
electro-weak model by introducing two spinors for any leptonic
family and quark generation and a scalar field with two components
with opposite hypercharge, responsible of spontaneous symmetry
breaking.

\end{titlepage}

\section{Introduction}
Once in 1916 General Relativity theory
appeared, by virtue of Albert Einstein's
genial recognition of the general space-time structure,
it arose the idea that all fundamental
interactions of Nature could admit a geometrical
interpretation, i.e. developed the so-called
``geometrical unification theories''.\\
In four space-time dimensions, the Riemmannian dynamics admits
enough degrees of freedom to describe only the gravitational
field, whereas all the others ones have to be regarded as
``matter'' which influences
the metric of the space-time by its energy-momentum.\\
As well known, a natural way to enlarge the
number of geometrical degrees of freedom consists of adding
extradimensions to the space-time and then using the new
available metric components to describe
other fundamental field of Nature.
Such a point of view was addressed first by
Kaluza and Klein,
who provided an independent geometrization of the
electromagnetic field via a 5-dimensional
space-time geometry;
it is just since them that this kind of approach
to geometrical unification of the fundamental
interactions acquired the name of
``Kaluza-Klein theories''.
It is worth noting how
the main achievement reached by Kaluza \cite{B15} consisted
of recognizing the multidimensional idea of unification,
while Klein \cite{B16} \cite{B17} had the merit to clarify how the
extradimension can acquire a precise physical
meaning under a suitable topology choice, i. e. the
so-called ``compactification of the dimension``.\\ 
After them the idea of multidimensionality acquired a physical plausibility 
and stimulated many works \cite{B35} \cite{Jo47};
however the Kaluza-Klein approach manifested all its powerful
capacity to represent Nature in a geometrical picture, only when
it was shown how the choice of an extradimensional compact
homogeneous space (for the proprieties of homogeneous spaces see
\cite{Lan}), allowed the geometrization of the non-Abelian
gauge theories, i.e. of the so-called Yang-Mills fields
\cite{DW64}-\cite{B14} (for a complete review see
\cite{OW97}); the leading idea of this success rely on the
correspondence between the isometry of the internal space
and the Lie algebra of the non-Abelian group.\\
In spite of the surprising performance of the Kaluza-Klein point
of view in geometrizing the bosonic component of the fundamental
interactions, it shows serious shortcomings when the attempt of
extending the procedure to the fermionic fields is faced (without
introducing supersymmetries in the scheme \cite{C78} \cite{Du86}),
thus in our work we introduce fermions as matter fields. However,
assuming suitable hypothesis for the dependence from
extra-coordinates, it is possible to
geometrize gauge connections \cite{M04}, thus we have to introduce just free spinors.\\
This assumption is also justified by the fact that it leads to the identification of extra-components of the fields momentum with gauge charges,
whose conservation thus is consequence of the invariance of the theory under extra-dimensional translations that can be interpreted as gauge transformations.\\
In our work, we consider a ``pure'' Kaluza-Klein approach, i.e.
extradimensions are spacelike and compact; however there are
alternative models, such as \emph{projective} theories \cite{L82}-\cite{Sch95}, in which they introduce not
physically real extracoordinates, and \emph{non compattified}
theories \cite{Wes90}-\cite{Wes96},
that assume noncompact extra-dimensional manifold with not necessary spacelike coordinates.\\
Here the only extension respect to canonical Kaluza-Klein theories is the introduction of a dependence of the extra-dimensional metric by four-dimensional coordinates by some scalar fields alpha, for which we get the dynamics.\\

The most important problem we face by extending the Kaluza-Klein approach to more than five dimensions is that a ground state direct product of a Minkosky space and a compattified one is not a solution of vacuum Einstein equations
\cite{Sa82} \cite{Ap83} \cite{bisAp83} \cite{Do84}. A possible solution to this feature is the introduction of suitable matter field, by which it is also possible to give an explanation for the breaking of general invariance with \emph{Spontaneous
Compactification} processes \cite{C77}-\cite{To84} (for a review \cite{Ba87}). They consist in the application of spontaneous symmetry breaking mechanism to space-time by introducing a field whose Lagrangian is
invariant under general coordinates transformations but it
is in a vacuum state which breaks symmetry,
maintaining just general four-dimensional invariance and
invariance under transformations on extra-space which reproduce on
fields gauge transformations.\\
However, as point out by Witten \cite{W81}, a Minkosky plus compact ground state can also be the minimum of some potential of a quantistic theory.\\
Other problems arise when we consider a non-abelian gauge group; in particular to reproduce the correct transformation law either for gauge bosons either for matter fields, and also to get the fields equations in the four-dimensional theory, every four-dimensional model has to be consider as a phenomenological theory built by an observer who cannot see extra-dimensions.\\
Moreover, free Dirac lagrangian dimensional splitting produces four-dimensional Dirac lagrangian with spinorial gauge connections plus adjunctive terms that can be eliminated with suitable assumptions for extra-dimensional spinorial connections \cite{M04}; this feature arises the problem of the physical interpretation to give to connections in a manifold along which we consider only translations.\\

In particular in our work we present, as first application of the
general case, the Kaluza-Klein theory in 5 dimensions which leads
to the geometrization of the electro-magnetic interaction, thus of a U(1)
gauge theory; we develop a theory in which the quantum
electrodynamics emerges in natural way when reducing all the
dynamical
variables to their 4-dimensional status.\\
Connected with this main result, we also face
the problem about the physical interpretation of the
scalar field associated to the $5-5$ component of the
metric tensor; indeed when we take $j_{55} = 1$,
in the 5-dimensional Lagrangian,
the presented theory reduces exactly to the
Einstein-Maxwell-Dirac dynamics, but the most general
case involves a scalar field whose dynamics do not
resemble any physical field.
We show that by
the introduction of a new scalar field,
linked to the
original one, and by a conformal transformation
operated on the four-dimensional metric,
it is possible to obtain
a term in the action
(as well as an energy-momentum tensor)
for the new scalar dynamics which looks
like that one of a Klein-Fock field.\\

Then, we demonstrate by a classical calculation the equivalence between
the fifth component of 5-momentum
and the electric charge and, following such a procedure, we also
obtain an estimate for
the length $L$ of the fifth dimension.  We reliably expect that this relation be valid in general and then
treating with charged particles we have to provide the
corresponding field with a phase dependence on
the fifth coordinate.\\

However it is Well known \cite{B82} that the introduction of a
spinorial field on a curved space-time, leads to define a new
covariant derivation made up of the ordinary one plus a new term,
which acts only on spinorial quantities via a specific connection
called just spinorial connection.\\
This new derivation is necessary to make possible the extension to
a curved space-time of the whole paradigm for the Dirac's algebra;
in particular the spinorial covariant derivative is constructed by
requiring that it anhilates the Dirac's matrices on curved
space-time.

Once defined the spinorial covariant derivative
for a five-dimensional space-time,
we show that if we assume a non-standard form for the connections we obtain the desired electrodynamics coupling.\\
The analysis of the pentadimensional case is concluded via considering the
modification of the spinorial part of the 4-dimensional action
produced by the conformal transformation on the 4-dimensional
metric tensor. More precisely, we show that, if we don't operate a
suitable conformal transformation on the spinorial field, then we
obtain a Dirac equation for $\psi$ and $\bar{\psi}$ which leads to
the non-conservation of the electric
charge.\\

The other application we consider is the geometrization of the electro-weak model, thus of a gauge group $SU(2)\otimes U(1)$; the problem that affect multidimensional theories which deal with electro-weak interaction is the so-called \emph{chirality problem} \cite{W81}
\cite{W83}, the problem to reproduce spinors with opposite chirality with different interaction proprieties. Here we overall it by assuming for different chirality states different dependence by extra-coordinates and we derive all standard model particles from just two spinors for every leptonic family and quark generation, thus obtaining a reduction of the number of matter fields than Standard Model.\\ Then, to reproduce Higgs bosons we introduce a scalar field subjected to Higgs potential and, to realize invariant fermionic mass terms, we predict it has two components with opposite hypercharge; however dimensional splitting produces adjunctive Higgs mass term $\approx 10^{19}GeV$ which will impose extremely accurate fine-tuning on the parameters of Higgs potential. Another sources of spontaneous symmetry breaking may be the fields responsible of spontaneous compattification process or alpha fields, which we show can be interpreted as Klein-Gordon fields.\\\\

In particular, the organization of all these features in the work is the following:
\begin{itemize} \item in the first chapter we deal with the general case:
in section 1 we achieve a generic gauge theory
geometrization in presence only of gauge bosons, obtained using an
action which is the n-dimensional generalization of Einstein-Hilbert
one, while in section 2 we introduce matter fields and show the
equivalence between gauge charges and extra-components of the
field momentum, whose conservation comes from the invariance of
the theory under extra-dimensional translations;
in section 3 we present the case of a free spinorial
n-dimensional field and we show how gauge connection for the
four-dimensional field associated has a geometric source; we also
introduce suitable extra-dimensional spinorial connections in
order to reproduce the
four-dimensional theory.\\
\item in the second chapter we consider the pentadimensional theory:
in section 1 we provide a brief review on the 5-dimensional
Kaluza-Klein theory, within which, the electromagnetic interaction
finds its natural framework of geometrization; in section 2 we
reproduce the classical calculation about the relationship between
the fifth component of the momentum and the electric charge and
give the classical estimate for the length $L$ of the
extra-dimension; in section 3 we introduce in our theory a
spinorial field as a matter field and show that it is possible to
obtain the gauge coupling of Q.E.D. if we assume a non-standard
form of spinorial connections and provide the spinorial field with
a phase dependence on the fifth coordinate, then we also give an
independent estimate of $L$ which agrees with the classical one;
in section 4 we define a new scalar field and operate a conformal
transformation on the four-dimensional metric in such a way that
the new field is a Klein-Fock one; in section 5 we study the
modification of the spinorial part of the 4-dimensional action due
to the conformal transformation operated on the 4-dimensional
metric and show that if we don't operate an appropriate
transformation on the spinorial field we get inevitably the
non-conservation of the electric charge; in section 6 we present a
cosmological implementation of the model by which we can explain
the compattification of the fifth dimension and we predict for
fundamental constants a dependence from space-time
extra-coordinates.\\
\item the third chapter is the application to a gauge group
$SU(2)\otimes U(1)$ in a space-time $V^{4}\otimes S^{1}\otimes
S^{2}$: in section 1 by introducing two spinors for every leptonic family and
quark generation we overall chirality problem and reproduce all standard model particle and, by imposing
after dimensional reduction the coincidence of our Lagrangian with
electro-weak plus Einstein-Hilbert ones, we also obtain an
estimate of extra-dimensions length;
in section 5 we consider a two components scalar field subjected to
a Higgs potential, whose four-dimensional reduction has the same lagrangian density as Higgs field
but two components with different hypercharge; moreover the
presence of extra-dimensions produces a mass term $\approx
10^{35}GeV$, so to be consistent with limits on Higgs' mass it is
necessary an extremely accurate fine-tuning on the parameter
$\mu^{2}$; in section 3 we demonstrate that it is possible to
interpret $\alpha$ fields, by a redefinition of them and a
conformal transformation on four dimensional metric, as interacting 
Klein-Gordon fields.
\end{itemize}

\newpage

\section{Kaluza-Klein theory in general case}

\subsection{Geometrization of a gauge theory bosonic component}
Let consider a n-dimensional space-time manifold $V^{4}\otimes B^{k}$, where $V^{4}$ is the ordinary four-dimensional manifold and $B^{k}$ is a k-dimensional ($k=n-4$) compact homogeneous space such that its Killing vectors reproduce the algebra of the gauge gauge we geometrize
\begin{equation}
\label{a1} \xi^{n}_{\bar{N}}\frac{\partial
\xi_{\bar{M}}^{m}}{\partial y^{n}}-\xi^{n}_{\bar{M}}\frac{\partial
\xi_{\bar{N}}^{m}}{\partial
y^{n}}=C^{\bar{P}}_{\bar{N}\bar{M}}\xi^{m}_{\bar{P}}
\end{equation}
where $C^{\bar{P}}_{\bar{N}\bar{M}}$ are gauge group's structures constants.\\

In general the space-time dimensionality is different from the gauge group one so
the number of Killing vector is different from n and we can impose just one of the reciprocity
conditions
\begin{equation}
\xi^{n}_{\bar{N}}\xi^{\bar{M}}_{n}=\delta^{\bar{M}}_{\bar{N}}\qquad
\xi^{n}_{\bar{M}}\xi^{\bar{M}}_{m}=\delta^{n}_{m},
\end{equation}
we will take the first one.\\

In the following we will consider the variables $x^{\mu}\quad
(\mu=0,\ldots,3)$ for the ordinary four-dimensional coordinates,
$y^{m}\quad (m=0,\ldots,k-1)$ for
extra-dimensional ones and $x^{A}\quad (A=0,\ldots,n-1)$ for both.\\

The theory we construct is invariant under the coordinates
transformations
\begin{equation}
\label{b1}
\left\{\begin{array}{c} x'^{\mu}=x'^{\mu}(x^{\nu})\\
y'^{m}=y^{m}+\omega^{\bar{N}}(x^{\nu})\xi^{m}_{\bar{N}}(y^{n})
\end{array}\right.,
\end{equation}
so we have general invariance under 4-dimensional transformations
and the invariance under translations along extra-dimensions, and
we write the metric as
\begin{equation}
\label{c1}
j_{AB}=\left(\begin{array}{c|c}g_{\mu\nu}(x^{\rho})+\gamma_{mn}(x^{\rho};y^{r})\xi^{m}_{\bar{M}}(y^{r})\xi^{n}_{\bar{N}}(y^{r})A^{\bar{M}}_{\mu}(x^{\rho})A^{\bar{N}}_{\nu}(x^{\rho})
& \gamma_{mn}(x^{\rho};y^{r})\xi^{m}_{\bar{M}}(y^{r})A^{\bar{M}}_{\mu}(x^{\rho}) \\\\
\hline\\
\gamma_{mn}(x^{\rho};y^{r})\xi^{n}_{\bar{N}}(y^{r})A^{\bar{N}}_{\nu}(x^{\rho})
& \gamma_{mn}(x^{\rho};y^{r})\end{array}\right)
\end{equation}
where we assume that $\gamma_{mn}$ also depends
from the ordinary space-time variables, by some scalar field
$\alpha$ that determine the structure of the space $B^{K}$
\begin{equation}
\label{d1}
\gamma_{mn}(x;y)=\eta_{mr}(x)\overline{\gamma}_{rn}(y)\qquad
\eta_{mn}(x)=\eta_{mn}\alpha^{m}\alpha^{n}
\end{equation}
where in the latter the indices m and n are not summed. Under the
transformations (\ref{b1}), with $\omega$ infinitesimal, the
extra-dimensional metric at the first order in $\omega$, using the
(\ref{c1}), does not change for the homogeneity of $B^{K}$
\begin{equation}
\label{e1}\overline{\gamma}'_{rn}(y)=\overline{\gamma}_{rn}(y),
\end{equation}
while the fields $A^{\bar{M}}_{\mu}$ behave like Abelian gauge's
bosons
\begin{equation}
\label{f1}A'^{\bar{M}}_{\mu}=\frac{\partial x'^{\nu}}{\partial
x^{\mu}}\bigg[A_{\nu}^{\bar{M}}-\frac{\partial\omega^{\bar{M}}}{\partial
x'^{\nu}}\bigg]
\end{equation}
and $g_{\mu\nu}$ like a four-dimensional tensor
\begin{equation}
g_{\mu\nu}(x)=\frac{\partial x'^{\rho}}{\partial
x^{\mu}}\frac{\partial x'^{\sigma}}{\partial
x^{\nu}}g'_{\rho\sigma}(x')
\end{equation}
so it can be identified with four-dimensional metric.\\
In order to interpret the fields $A^{\bar{M}}_{\mu}$ as a
generic gauge theory's bosons it is necessary to consider the
way an observer who perceives a four-dimensional space-time
reveals a transformation on extra-dimensions, which are
compattified to such distance he cannot see them. In fact, the
variation of the component $j_{m\mu}$ under the second of
(\ref{b1}) is \cite{Lu78} at the first order in $\omega$
\begin{equation}
\delta
\label{g1}j_{m\mu}(x;y)=j'_{m\mu}(x;y)-j_{m\mu}(x;y)=\gamma_{mn}(x;y)\xi^{n}_{\bar{M}}(y)\bigg(C^{\bar{M}}_{\bar{P}\bar{Q}}\omega^{\bar{Q}}(x)A^{\bar{P}}_{\mu}(x)+
\frac{\partial \omega^{\bar{M}}(x)}{\partial x^{\mu}}\bigg)
\end{equation}
where the variation is a consequence of both the variations of
$A^{\bar{M}}_{\mu}$ and $\xi^{m}_{\bar{M}}$; but Killing vectors
are defined on extra-dimensional manifold and so neither them
neither their changing are observable. Consequently, a
four-dimensional observer interprets (\ref{g1}) as due only to the
variations of the fields $A^{\bar{M}}_{\mu}$
\begin{equation}
\delta j_{m\mu}(x;y)=\gamma_{mn}(x;y)\xi^{n}_{\bar{M}}(y)\delta
A^{\bar{M}}_{\mu}
\end{equation}
and, thus, at the first order in $\omega$
\begin{equation}
A_{\mu}^{\prime\bar{M}}(x)=A_{\mu}^{\bar{M}}(x)+C^{\bar{M}}_{\bar{P}\bar{Q}}\omega^{\bar{Q}}(x)A^{\bar{P}}_{\mu}(x)+
\frac{\partial \omega^{\bar{M}}(x)}{\partial x^{\mu}}.
\end{equation}
Therefore, \emph{from the point of view of an observer who cannot
see extra-dimensions, the fields $A^{\bar{M}}_{\mu}$ behave under
translations on the extra-coordinates like gauge bosons under
gauge transformations}, where the group's structures
constants are those which come from Killing vector's algebra.\\
\\
To invert the matrix $j_{AB}$ we define a new dual space base
\begin{equation}
d\widetilde{x}^{\mu}=dx^{\mu}\qquad
d\widetilde{y}^{m}=dy^{m}+\xi^{m}_{\bar{M}}A^{\bar{M}}_{\mu}dx^{\mu}
\end{equation}
and the new metric is
\begin{equation}
\label{h1}\widetilde{j}_{AB}=\left(\begin{array}{c|c}g_{\mu\nu} & 0 \\
\hline 0 & \gamma_{mn} \end{array}\right)
\end{equation}
which can easily be inverted
\begin{equation}
\widetilde{j}^{AB}=\left(\begin{array}{c|c}g^{\mu\nu} & 0 \\
\hline 0 & \gamma^{mn} \end{array}\right)
\end{equation}
and we have
\begin{equation}
\label{i1}j^{AB}=\left(\begin{array}{c|c}g^{\mu\nu}(x^{\rho})
& \-g^{\rho\mu}\xi^{n}_{\bar{M}}(y^{r})A^{\bar{M}}_{\rho}(x^{\sigma}) \\\\
\hline\\
\-g^{\rho\nu}\xi^{m}_{\bar{M}}(y^{r})A^{\bar{M}}_{\rho}(x^{\sigma})
&
\gamma^{mn}(x^{\rho};y^{r})+\xi^{m}_{\bar{M}}(y^{r})\xi^{n}_{\bar{N}}(y^{r})A^{\bar{M}}_{\mu}(x^{\rho})A^{\bar{N}}_{\nu}(x^{\rho})\end{array}\right).
\end{equation}
Moreover, we get
\begin{equation}
j\equiv det(j_{AB})=det(\widetilde{j}_{AB})
\end{equation}
and so, from (\ref{h1})
\begin{equation}
\label{j1}det(j_{AB})=\gamma g
\end{equation}
where g is the determinant of $g_{\mu\nu}$ and $\gamma$ of
$\gamma_{mn}$.\\\\
We want to derive the dynamic of the fields contained in the
metric from a variational principle, so we consider the action
which is the n-dimensional extension of the Einstein-Hilbert one
\begin{equation}
\label{k1}S=-\frac{c^{4}}{16\pi G_{(n)}}\int{}^n\!R\sqrt{-j}
d^{4}xd^{k}y
\end{equation}
and we use the tetradic formalism to calculate it and to split the
curvature into pure four-dimensional terms and terms which descend
from the presence of adjunctive dimensions and which describe
$A^{\bar{M}}_{\mu}$ and $\alpha^{m}$ dynamics.\\
From the relation of tetradic vectors $e^{(A)}_{A}$ with the
metric
\begin{equation}
\label{l1}j_{AB}=\eta_{(A)(B)}e^{(A)}_{A}e^{(B)}_{B}
\end{equation}
we have the following expressions
\begin{equation}
\label{m1}\left\{\begin{array}{c} g_{\mu\nu}=\eta_{(\mu)(\nu)}e_{\mu}^{(\mu)}e_{\nu}^{(\nu)}\\
e_{\mu}^{(m)}=e^{(m)}_{m}\xi^{m}_{\bar{M}}A^{\bar{M}}_{\mu}\\
e_{m}^{(\mu)}=0\\\gamma_{mn}=\eta_{(m)(n)}e_{m}^{(m)}e_{n}^{(n)}\end{array}\right.,
\end{equation}
and using the reciprocity
conditions $e^{A}_{(A)}e^{(A)}_{B}=\delta^{A}_{B}\quad
e^{A}_{(B)}e^{(A)}_{A}=\delta^{(A)}_{(B)}$
\begin{equation}
\label{n1}\left\{\begin{array}{c} g^{\mu\nu}=\eta^{(\mu)(\nu)}e^{\mu}_{(\mu)}e^{\nu}_{(\nu)}\\
e^{m}_{(\mu)}=-e^{\mu}_{\mu}\xi^{m}_{\bar{M}}A^{\bar{M}}_{\mu}\\e^{\mu}_{(m)}=0\\
\gamma^{mn}=\eta^{(m)(n)}e^{m}_{(m)}e^{n}_{(n)}\end{array}
\right..
\end{equation}
From (\ref{a1}), (\ref{l1}) and Killing equation
\begin{equation}
\frac{\partial\xi^{r}_{\bar{M}}}{\partial
y^{n}}\gamma_{mr}+\frac{\partial\xi^{r}_{\bar{M}}}{\partial
y^{m}}\gamma_{rn}+\xi^{r}_{\bar{M}}\frac{\partial
\gamma_{mn}}{\partial y^{r}}=0,
\end{equation}
we get two relations that will be useful in the following
\begin{equation}
\label{o1}\frac{\partial\xi^{r}_{\bar{M}}}{\partial
y^{n}}=-\xi^{s}_{\bar{M}}e^{r}_{(n)}\frac{\partial
e^{(n)}_{n}}{\partial y^{s}}.
\end{equation}
\begin{equation}
\label{p1}\partial_{p}e^{(r)}_{n}-\partial_{n}e^{(r)}_{p}=C^{\bar{P}}_{\bar{Q}\bar{M}}\xi^{s}_{\bar{P}}\xi_{p}^{\bar{Q}}\xi_{n}^{\bar{M}}
e^{(r)}_{s}.
\end{equation}
\\
We can get ${}^n\!R$ from its relations with Ricci
rotation coefficients that we obtain from
the anolonomy tensor.

We carry out the calculation in a local Lorentz frame, in
which
\begin{equation}
\label{u1}e^{\bar{\mu}}_{\mu}=\delta^{\bar{\mu}}_{\mu}
\end{equation}
and at the end we will restore the general invariance by
substituting all the ordinary derivative respect to the variables
$x^{\mu}$ with the covariant ones, respect to the same
variables.\\
Using (\ref{p1}) and (\ref{u1}) we have for the
anolonomy tensor
\begin{equation}
\left\{\begin{array}{c}\lambda_{(m)(\mu)(\nu)}=
\eta_{(m)(n)}\delta^{\mu}_{(\mu)}\delta^{\nu}_{(\nu)}e^{(n)}_{r}
\xi^{r}_{\bar{M}}F^{\bar{M}}_{\mu\nu}\\\\
\lambda_{(m)(n)(\mu)}=\eta_{(m)(r)}e^{n}_{(n)}\delta^{\mu}_{(\mu)}
(\partial_{\mu}e^{(r)}_{n}-2A^{\bar{M}}_{\mu}e^{(r)}_{s}\xi^{s}_{\bar{P}}
\xi^{\bar{Q}}_{n}C^{\bar{P}}_{\bar{Q}\bar{M}})\\\\
\lambda_{(m)(n)(p)}=\eta_{(m)(r)}
e^{n}_{(n)}e^{p}_{(p)}e^{(r)}_{s}C^{\bar{P}}_{\bar{Q}\bar{M}}
\xi^{s}_{\bar{P}}\xi^{\bar{Q}}_{p}\xi^{\bar{M}}_{n}
\end{array}\right.
\end{equation}
where
\begin{equation}
F^{\bar{M}}_{\mu\nu}=\partial_{\nu}A^{\bar{M}}_{\mu}-\partial_{\mu}A^{\bar{M}}_{\nu}
+C^{\bar{M}}_{\bar{N}\bar{P}}A^{\bar{N}}_{\mu}A^{\bar{P}}_{\nu}
\end{equation}
and for Ricci rotation coefficients
\begin{equation}
\left\{\begin{array}{c}R_{(\mu)(\nu)(m)}=-\frac{1}{2}\eta_{(m)(n)}\delta^{\mu}_{(\mu)}\delta^{\nu}_{(\nu)}e^{(n)}_{r}\xi^{r}_{\bar{M}}F^{\bar{M}}_{\mu\nu}\\\\
R_{(m)(\mu)(\nu)}=\frac{1}{2}\eta_{(m)(n)}\delta^{\mu}_{(\mu)}\delta^{\nu}_{(\nu)}e^{(n)}_{r}\xi^{r}_{\bar{M}}F^{\bar{M}}_{\mu\nu}\\\\
R_{(\mu)(m)(n)}=\lambda_{\{(m)(n)\}(\mu)}\\\\
R_{(m)(n)(\mu)}=\lambda_{[(m)(n)](\mu)}\\\\
R_{(m)(n)(p)}=\frac{1}{2}[\lambda_{(m)(n)(p)}+\lambda_{(n)(p)(m)}+\lambda_{(p)(n)(m)}]
\end{array}\right.
\end{equation}
from which, by means of the hypothesi that
$C^{\bar{P}}_{\bar{M}\bar{N}}$ are totally antisymmetric,
(\ref{d1}) and (\ref{o1})
\begin{equation}\label{curvsplit}
{}^n\!R=R-\frac{1}{4}\gamma_{rs}\xi^{r}_{\bar{M}}\xi^{s}_{\bar{N}}F^{\bar{M}}_{\mu\nu}F^{\bar{N}}_{\rho\sigma}g^{\mu\rho}g^{\nu\sigma}-
2g^{\mu\nu}\sum_{n=1}^{k}\frac{\nabla_{\mu}\partial_{\nu}\alpha^{n}}{\alpha^{n}}-g^{\mu\nu}\sum_{n\neq
m=1}^{k}\frac{\partial_{\mu}\alpha^{m}}{\alpha^{m}}\frac{\partial_{\nu}\alpha^{n}}{\alpha^{n}}+R_{N}
\end{equation}
where $R$ is the four-dimensional curvature and $R_{N}$ the curvature term
\begin{eqnarray*}
R_{N}=\frac{1}{2}\gamma^{ns}\xi^{\bar{Q}}_{n}\xi^{\bar{T}}_{s}C^{\bar{P}}_{\bar{R}\bar{Q}}C^{\bar{R}}_{\bar{T}\bar{P}}
-\frac{1}{4}\gamma^{mr}\gamma^{ns}\gamma_{tu}\xi^{\bar{M}}_{m}\xi^{\bar{T}}_{r}\xi^{\bar{Q}}_{n}\xi^{\bar{S}}_{s}\xi^{t}_{\bar{P}}
\xi^{u}_{\bar{R}}C^{\bar{P}}_{\bar{Q}\bar{M}}C^{\bar{R}}_{\bar{T}\bar{S}}
\end{eqnarray*}
and just contains interactions among the fields $\alpha$.\\
So the action (\ref{k1}) is
\begin{eqnarray*}
S=-\frac{c^{3}}{16\pi G_{(n)}}\int_{V^{4}\otimes B^{K}}
\sqrt{-\gamma}\sqrt{-g}\bigg[R-\frac{1}{4}\gamma_{rs}\xi^{r}_{\bar{M}}\xi^{s}_{\bar{N}}F^{\bar{M}}_{\mu\nu}F^{\bar{N}}_{\rho\sigma}g^{\mu\rho}g^{\nu\sigma}-
\\-2g^{\mu\nu}\sum_{n=1}^{k}\frac{\nabla_{\mu}\partial_{\nu}\alpha^{n}}{\alpha^{n}}-g^{\mu\nu}\sum_{n\neq
m=1}^{k}\frac{\partial_{\mu}\alpha^{m}}{\alpha^{m}}\frac{\partial_{\nu}\alpha^{n}}{\alpha^{n}}+R_{N}\bigg]d^{4}xd^{k}y;
\end{eqnarray*}
and assuming that given a Killing vector $\xi^{r}_{\bar{M}}$ all
$\alpha^{r}$ are equal for vector's r-components which are not
zero and are defined $\alpha^{\bar{M}}$, we have
\begin{eqnarray*}
S=-\frac{c^{3}}{16\pi G}\int_{V^{4}}
\sqrt{-g}\bigg[R-\frac{1}{4}\eta_{\bar{M}\bar{N}}\alpha^{\bar{M}}\alpha^{\bar{N}}F^{\bar{M}}_{\mu\nu}F^{\bar{N}}_{\rho\sigma}g^{\mu\rho}g^{\nu\sigma}-
\end{eqnarray*}
\begin{equation}
\label{azsplit}-2g^{\mu\nu}\sum_{n=1}^{k}\frac{\nabla_{\mu}\partial_{\nu}\alpha^{n}}{\alpha^{n}}-g^{\mu\nu}\sum_{n\neq
m=1}^{k}\frac{\partial_{\mu}\alpha^{m}}{\alpha^{m}}\frac{\partial_{\nu}\alpha^{n}}{\alpha^{n}}+R'_{N}\bigg]d^{4}x
\end{equation}
with the following positions
\begin{equation}
G=\frac{G_{(n)}}{V^{K}}
\end{equation}
\begin{equation}
\int_{B^{K}}\sqrt{-\gamma}[\overline{\gamma}_{rs}\xi^{r}_{\bar{M}}\xi^{s}_{\bar{N}}]d^{k}y=V^{K}\eta_{\bar{M}\bar{N}}
\end{equation}
\begin{equation}
R'_{N}=\frac{1}{V^{K}}\int_{B^{K}}\sqrt{-\gamma}R_{N}d^{k}y.
\end{equation}
where $V^{K}$ is $B^{K}$ space's volume.\\
Thus we get \emph{the geometrization of a generic gauge theory and
its unification with gravity, in the sense that both
Einstein-Hilbert action and Yang-Mills action derive from
dimensional splitting of n-dimensional curvature}.\\
So gauge bosons can be interpreted like geometrical terms; the
next steps are to introduce matter and to geometrize its coupling
with them.

\subsection{Momentum extra-components as gauge charges}
Let introduce in empty space-time $M^{4}\otimes
B^{K}$ some matter fields
$\varphi_{r}$ whose dynamic is described by a lagrangian density
\begin{equation}
\Lambda=\Lambda(\varphi_{r};\partial_{A}\varphi_{r})
\end{equation}
which is invariant under the infinitesimal coordinates
transformations
\begin{equation}
\left\{\begin{array}{c} x'^{\mu}=x^{\mu}+\delta\omega^{\mu}\\
y'^{m}=y^{m}+\delta\omega^{\bar{P}}\xi^{m}_{\bar{P}}
\end{array}\right.
\end{equation}
that can be rewrite as
\begin{equation}
x'^{A}=x^{A}+\delta\omega^{\bar{B}}u^{A}_{\bar{B}}
\end{equation}
where
$\delta\omega^{\bar{A}}=(\delta\omega^{\bar{\mu}};\delta\omega^{\bar{P}})$
and
\begin{equation}
u^{A}_{\bar{\mu}}=(\delta^{\mu}_{\bar{\mu}};0)\qquad
u^{A}_{\bar{M}}=(0;\xi^{m}_{\bar{M}}).
\end{equation}
Let consider the invariance under global transformations, thus we
choice $\delta\omega^{\bar{A}}$ constant, so we get for fields
transformations
\begin{equation}
\varphi'_{r}=\varphi_{r}+\delta\varphi_{r},\qquad
\delta\varphi_{r}=\partial_{A}\varphi_{r}u^{A}_{\bar{B}}\delta\omega^{\bar{B}}
\end{equation}
and for lagrangian density
\begin{eqnarray*}
\delta\Lambda=(\partial_{A}\Lambda)
u^{A}_{\bar{B}}\delta\omega^{\bar{B}}=\frac{
\partial\Lambda}{\partial\varphi_{r}}\delta\varphi_{r}+\frac{
\partial\Lambda}{\partial(\partial_{A}\varphi_{r})}\delta(\partial_{A}\varphi_{r})=\\\\
=\partial_{A}\bigg(\frac{
\partial\Lambda}{\partial(\partial_{A}\varphi_{r})}\delta\varphi_{r}\bigg)-\bigg[\partial_{A}\bigg(\frac{
\partial\Lambda}{\partial(\partial_{A}\varphi_{r})}\bigg)-\frac{
\partial\Lambda}{\partial\varphi_{r}}\bigg]\delta\varphi_{r}
\end{eqnarray*}
\\
from which, using Euler-Lagrange equations in a curved space
\begin{equation}
\nabla_{A}\bigg(\frac{
\partial\Lambda}{\partial(\partial_{A}\varphi_{r})}\bigg)-\frac{
\partial\Lambda}{\partial\varphi_{r}}=0
\end{equation}
and the condition that the connection is metric for $\gamma_{mn}$
\begin{equation}
\nabla_{r}\gamma_{mn}=0
\end{equation}
which lead to
\begin{equation}
\nabla_{A}u^{A}_{\bar{B}}=0,
\end{equation}
we have
\begin{equation}
\label{a2}\nabla_{A}\bigg(\frac{
\partial\Lambda}{\partial(\partial_{A}\varphi_{r})}(\partial_{C}\varphi_{r})u^{C}_{\bar{B}}-\Lambda
u^{A}_{\bar{B}}\bigg)=0.
\end{equation}
The (\ref{a2}) is a continuity equation and the associated
conserved quantities are n-dimensional components of fields
momentum \cite{B8}
\begin{equation}
\label{b2}P_{\bar{A}}=\int_{E^{3}\otimes B^{K}}
\sqrt{-\gamma}[\Pi_{r}(\partial_{B}\varphi_{r})u^{B}_{\bar{A}}-\Lambda
u^{0}_{\bar{A}}]d^{3}xd^{k}y
\end{equation}
\\
where $E^{3}$ is euclidean three-dimensional space and $\Pi_{r}$
the fields conjugated to $\varphi_{r}$
\begin{equation}
\Pi_{r}=\frac{
\partial\Lambda}{\partial(\partial_{t}\varphi_{r})}.
\end{equation}
Thus if we assume that $\varphi_{r}$ dependence from
extra-coordinates is of the type
\begin{equation}
\varphi_{r}=\frac{1}{\sqrt{V^{K}}}e^{-i\tau_{rs}(y^{m})}\phi_{s}(x^{\mu})\qquad
\Pi_{r}=\frac{1}{\sqrt{V^{K}}}\pi_{s}(x^{\mu})e^{i\tau_{sr}(y^{m})}\label{c2}
\end{equation}
we have
\begin{eqnarray}
Q_{\mu}=P_{\mu}=\int_{E^{3}}[\pi_{r}\partial_{\mu}\phi_{r}-\Lambda\delta^{0}_{\mu}]d^{3}x\\
\label{e2}Q_{\bar{M}}={-i}\int_{E^{3}\otimes
B^{K}}\frac{\sqrt{-\gamma}}{V^{K}}[\pi_{r}\xi^{n}_{\bar{M}}\partial_{n}\tau_{rs}(y^{m})\phi_{s}]
d^{3}xd^{K}y
\end{eqnarray}
the first terms are the ordinary four-dimensional components of
fields momentum, which are therefore conserved, while to interpret
the other ones like conserved gauge charges we put \cite{M04}
\begin{equation}
\label{d2}\tau_{rs}(y^{m})=T_{\bar{P}rs}\lambda^{\bar{P}}_{\bar{Q}}\Theta^{\bar{Q}}(y^{m})
\end{equation}
with $T_{\bar{P}}$ the generators of the gauge group that acts on
the fields $\phi_{s}$, $\Theta^{\bar{Q}}$ functions of the
extra-dimensional variables expandable in generalized Fourier
series \footnote{We stress how the functions $\Theta^{P}$ can not be pure scalars, otherwise the matrix $(\lambda^{-1})^{P}_{Q}$ vanishes identically. On the other hand they have to be invariants under extra-coordinates transformations (isometries); thus the natural choice for $\Theta^{P}$ is to take scalar density of weight $\frac{1}{2}$, i.e. $\Theta^{P}=\sqrt{\gamma}\phi^{P}(y^{l})$, being $\phi^{P}$ non-constant scalars.} and the constant matrix $\lambda$ such that
\begin{equation}
\label{f2}(\lambda^{-1})^{\bar{P}}_{\bar{Q}}=\frac{1}{V^{K}}\int_{B^{K}}
\sqrt{-\gamma}\Big(\xi^{m}_{\bar{Q}}\partial_{m}\Theta^{\bar{P}}\Big)d^{K}y.
\end{equation}
In fact from (\ref{c2}), (\ref{d2}) and (\ref{f2}) we can rewrite
(\ref{e2}) as
\begin{equation}
Q_{\bar{M}}=-i\int_{E^{3}}(\pi_{r}T_{\bar{M}rs}\phi_{s})d^{3}x,
\end{equation}
so \emph{we can interpret gauge charges as extra-dimensional
components of fields momentum} and their conservation comes from
the invariance of the theory under translations along
extra-dimensional coordinates.\\
This suggests to identify gauge transformations with
extra-dimensional translations, but from (\ref{c2}) the
transformation law for the fields is
\begin{equation}
\label{g2}\phi'_{r}=\phi_{r}+i\delta\omega^{\bar{Q}}\lambda^{\bar{N}}_{\bar{Q}}
\xi^{m}_{\bar{N}}\partial_{m}\Theta^{\bar{P}}T_{\bar{P}rs}\phi_{s}
\end{equation}
which coincides with gauge transformation law only if
\begin{equation}
\xi^{m}_{\bar{N}}\partial_{m}\Theta^{\bar{P}}=\delta^{\bar{P}}_{\bar{N}}
\end{equation}
and the last equation is equivalent to the vanishing of all
structures constants; thus the equivalence between gauge
transformation and extra-dimensional translations is immediate
only in the Abelian case. \\
Nevertheless, we have to consider that an observer, because of
compattification, cannot see extra-dimensions, so he considers as
the same physical state states with equal ordinary
four-dimensional coordinates and different extra-dimensional ones
and, hence, the result of the measurement of a physical quantity
on an eigenstate of position is obtained by an integration on
the adjunctive
variables.\\
Therefore, the fields transformation law viewed by a
four-dimensional observer is
\begin{equation}
\phi'_{r}=\frac{1}{V^{K}}\int
\sqrt{-\gamma}(\phi_{r}+i\delta\omega^{\bar{Q}}\lambda^{\bar{N}}_{\bar{Q}}
\xi^{m}_{\bar{N}}\partial_{m}\Theta^{\bar{P}}T_{\bar{P}rs}\phi_{s})d^{k}y
=\phi_{r}+i\delta\omega^{\bar{Q}}T_{\bar{Q}rs}\phi_{s}.
\end{equation}
which is the correct law for fields under gauge transformations.\\
We, thus, interpret \emph{gauge transformations as the result of
translations on the extra-dimensions and of their
compattification}.\\
\\
The same considerations stand for equations of fields motion in
order to eliminate the dependence from extra-coordinates which
obviously are not present in four-dimensional theory. In this
sense we remaind to next work the check that from n-dimensional
Einstein equations in the vacuum, which are the equations one gets
from (\ref{k1}), by an integration on extra-dimensional
variables we obtain four-dimensional Einstein equations in
presence of gauge bosons fields and of $\alpha$ fields.

\subsection{N-dimensional spinorial field}
We introduce matter as spinorial fields and assume a lagrangian
density which is the n-dimensional extension of Dirac one
\begin{equation}
\label{a3}\Lambda=\frac{i\hbar c}{2}
[D_{(A)}\bar{\Psi}\gamma^{(A)}\Psi-\bar{\Psi}\gamma^{(A)}D_{(A)}\Psi];
\end{equation}
where $\gamma^{(A)}$ are n matrix that satisfy the conditions
of Dirac algebra
\begin{equation}
\left\{\begin{array}{c}[\gamma^{(A)};\gamma^{(B)}]=2I\eta^{(A)(B)}
\\\partial_{A}\gamma^{(B)}=0\\(\gamma^{(A)})^{\dag}=\gamma^{(0)}\gamma
^{(A)}\gamma^{(0)}\end{array}\right.
\end{equation}
and
\begin{equation}
D_{(A)}\Psi=\partial_{(A)}\Psi-\Gamma_{(A)}\Psi\qquad
D_{(A)}\bar{\Psi}=\partial_{(A)}\bar{\Psi}+\bar{\Psi}\Gamma_{(A)}.
\end{equation}
In the following we assume that $\gamma^{(\mu)}$ and
$\Gamma_{(\mu)}$ are equal to the four-dimensional ones.\\
Now, we can carry out the dimensional reduction of the action
which we get from (\ref{a3})
\begin{eqnarray*}
S=\frac{ic\hbar}{2}\int\sqrt{\gamma}\sqrt{-g}
[(\partial_{(\mu)}+\Gamma_{(\mu)})\bar{\Psi}\gamma^{(\mu)}\Psi-
\bar{\Psi}\gamma^{(\mu)}(\partial_{(\mu)}-\Gamma_{(\mu)})\Psi+\\+
(\partial_{(m)}+\Gamma_{(m)})\bar{\Psi}\gamma^{(m)}\Psi-\bar{\Psi}\gamma^{(m)}(\partial_{(m)}
-\Gamma_{(m)})\Psi]d^{4}xd^{k}y
\end{eqnarray*}
and, using (\ref{c2}),(\ref{d2}) and (\ref{f2}), by the integration on
extra-coordinates we have
\begin{eqnarray*}
S=\frac{ic\hbar}{2}\int
[(\partial_{(\mu)}-ie^{\mu}_{(\mu)}A_{\mu}^{\bar{M}}T_{\bar{M}}+\Gamma_{(\mu)})\bar{\psi}
\gamma^{(\mu)}\psi-\bar{\psi}\gamma^{(\mu)}(\partial_{(\mu)}+ie^{\mu}_{(\mu)}A_{\mu}^{\bar{M}}T_{\bar{M}}
-\Gamma_{(\mu)})\psi+
\end{eqnarray*}
\begin{equation}
+(-\widetilde{\Gamma}_{(m)}+\Gamma_{(m)})\bar{\psi}\gamma^{(m)}\psi-
\bar{\Psi}\gamma^{(m)}(\widetilde{\Gamma}_{(m)}
-\Gamma_{(m)})\psi]\sqrt{-g}d^{4}x
\end{equation}
where $\psi=\psi(x^{\mu})$ and
\begin{equation}
\widetilde{\Gamma}_{(m)}=\frac{i}{V^{K}}\lambda^{\bar{P}}_{\bar{Q}}T_{\bar{P}}\int_{B^{K}}
(\sqrt{-\gamma}e^{m}_{(m)}\partial_{m}\Theta^{\bar{Q}})d^{K}y.
\end{equation}
Thus, the interaction terms between gauge bosons and spinorial
fields descends naturally by the splitting, but the theory also
predict adjunctive terms that can be eliminated imposing
\begin{equation}
\Gamma_{(m)}=\widetilde{\Gamma}_{(m)};\label{spconn}
\end{equation}
making this choice we are considering \emph{a n-dimensional
space-time which is not a Riemannian manifold}, but $V^{4}$ is
still riemannian, because for the four-dimensional connection we
take the standard form \cite{B82}
\begin{equation}
\Gamma_{(\mu)}=-\frac{1}{4}\gamma^{(\nu)}\nabla_{(\mu)}\gamma_{(\nu)}.
\end{equation}
\\
The problem we have taking spinorial extra-connections not zero concerns
physical interpretation of them, because a spinor does not change
under translations, which are the transformations we consider in
extra-dimensional space, we expect the connections to be zero. But
there is also to consider that the dependence of fields by the
coordinates is very different between the ordinary and the extra
ones (\ref{c2}), so our physical interpretation about
connections could fail when we face extra-dimensions.\\
However, taking zero extra-connections would produce the ordinary
four-dimensional theory plus terms which describe free currents.

\section{Kaluza-Klein theory in 5-dimensions (geometrization of the gauge group U(1))}

\subsection{Kaluza-Klein theory}

In the original Kaluza-Klein theory is postulated as space-time a
5-dimensional (smooth) manifold $V^5$ of class $C^\infty$, which
in order to reproduce the algebra of a U(1) group is the direct
product between a generic 4-dimensional manifold and a circle of
radius $\rho$, i.e. $V^4\times S^1$; the analysis relies on the
hypothesis (cylindricity condition) that there is, in the problem,
no dependence on the fifth coordinate, that is

\begin{equation}j_{AB}=j_{AB}(x^\mu)
\qquad \{\mu=0,1,2,3\}
\, ,
\label{5}
\end{equation}

indeed the same restriction must hold  for all
the observable present in the theory.\\
It is worth noting how, the requirement to have a compact fifth
dimension implies that the  metric be periodic  in the
corresponding coordinate; hence the independence on $x^5$ has to
be regarded as a zero-order cutoff of a Fourier expansion of all
physical quantities.\\

According to (\ref{c1}),  the 15 components of the 5-dimensional metric
and of its inverse can be recasted in terms of the following 4-dimensional
scalar, vector and tensor quantities

\begin{eqnarray}j_{AB}=\left(\begin{array}{cc}
g_{\mu\nu}+
\Phi^2A_\mu A_\nu & ekA_\mu \\ ekA_\mu & \Phi^2
\end{array}\right)\end{eqnarray}

\begin{eqnarray}j^{AB}=\left(\begin{array}{cc}
g^{\mu\nu} & -ekA^\mu \\ -ekA^\mu & \frac{1}{\Phi^2}+e^2k^2A_\rho
A^\rho
\end{array}\right)\label{a^(-1)2}\end{eqnarray}                                     

where $A_\mu$ transforms like an abelian gauge boson,
$e$ will be identified with the electric charge and $k$
denotes  a constant implied by dimensional
considerations, $g_{\mu\nu}$ is the four-dimensional metric and $\Phi$
a scalar field.\\

To get the field equations
associated to our assumptions,
we adopt
a variational principle by taking as
Lagrangian density $^5\!\Lambda$ the
pentadimensional extension of the
Einstein-Hilbert one, i.e.

\begin{equation}^5\!\Lambda=-\frac{c^4}
{16\pi G_5}{}^5\!R \label{19}                                
\end{equation}
and the action
\begin{equation}^5\!S=-\frac{c^3}
{16\pi G_5}\int\sqrt{-j}\,\,{}^5\!
R\,
d{}^5\!\Omega
,\, .
\label{20}
\end{equation}

Being

\begin{equation} j\equiv det((j_{AB}))=
det((\widetilde{j}_{AB}))\equiv
\widetilde{j}=\Phi\sqrt{-g}
\, ,
\label{21}
\end{equation}

from (\ref{curvsplit}) we get for the quadridimensional action
\begin{equation}
S=-\frac{c^3}{16\pi G}\int\sqrt{-g}
[\Phi R+2\nabla_\mu\partial^\mu\Phi
+\frac{e^2k^2}{4}\Phi^3F_{\mu\nu}F^{\mu\nu}]
\,d\Omega
\,;
\label{36}                                                
\end{equation}
we observe that in the pentadimensional case the curvature terms vanish.\\

By varying such action with respect to
the fields $g^{\alpha\beta}$, $\Phi$
and $A_\mu$ one obtains a set of 4-dimensional equations of the form

\begin{eqnarray}
\left\{\begin{array}{ccc}\Phi G^{\alpha\beta}+
\nabla^\beta(\partial^\alpha\Phi)-
g^{\alpha\beta}\nabla_\gamma
(\partial^\gamma\Phi)+
\frac{e^2k^2}{2}\Phi^3
[F^\alpha_{\phantom{\alpha}\mu}F^{\beta\mu}-
\frac{1}{4}g^{\alpha\beta}
F_{\mu\nu}F^{\mu\nu}]=0
{}\\{}
R+\frac{3e^2k^2}{4}\Phi^2
F_{\mu\nu}F^{\mu\nu}=0
\qquad \qquad \qquad \qquad
\qquad \qquad \qquad \qquad \qquad
\qquad\,{}\\{} \nabla_\nu
(\Phi^3 F^{\nu\mu})=0   \qquad \qquad
\qquad \qquad \qquad \qquad \qquad
\qquad \qquad \qquad \qquad\quad\,\,\, .
\label{37}
\end{array}\right.\end{eqnarray}
from this system, as soon as we set
$\Phi=1$, we find the usual
Einstein-Maxwell equations,
but also the unphysical compatibility
condition

\begin{equation} F_{\mu\nu}F^{\mu\nu}=0.
\,
\label{38}
\end{equation}

To avoid such an inconsistence of the theory, we are lead
to impose the condition $\Phi=1$ before varying
the action; in this way we cannot variate respect to $\Phi$ so that the second equation in the (\ref{37}) and, thus, the unphysical condition does not outcome.\\

\subsection{Fifth Component of the Momentum
and the Electric Charge}

At this point, using the condition $\Phi=1$, we point out that even by a classic
calculation we can get the equivalence existing
between the fifth component of the
pentamomentum of a particle and the
electric charge of the same.\\
We can start assuming an incoherent
dust for which the particles move
along the geodesics.\\
For this reason we assume as action
\begin{eqnarray}
    S=-mc\int\sqrt{j_{AB}u^Au^B}\,ds
     \label{68.5}
\end{eqnarray}
which leads to the equations
\begin{equation}
  u^B\nabla_B u_A=0.  \label{70}
\end{equation}
Putting $A=5$, because of the
cylindricity condition, one obtains
\begin{equation}
  u^\nu\nabla_\nu u_5=0\qquad
  \Rightarrow\qquad u_5=cost
  \label{71}
\end{equation}
Now, we can calculate the constant k simply by imposing that from the 
equation $(\ref{37})_1$, in which
we put $\Phi=1$, we get 
Einstein's equations
\begin{equation}
  G_{\alpha\beta}=-\frac{8\pi G}{c^4}
  \frac{1}{4\pi}[F_{\alpha\rho}
  F_{\beta}^{\phantom{\beta}\rho}-
  \frac{1}{4}g_{\alpha\beta}
  F_{\mu\nu}F^{\mu\nu}].  \label{72}                                 
\end{equation}
The condition needed is just 
\begin{equation}
  k\equiv\sqrt{4G}/ec^2  \label{73}                     
\end{equation}
and in what follows we will assume it.\\
Putting $A=\mu$ in (\ref{70}), neglecting
the terms in which the quantities
$A_\nu$ and $\nabla_\nu A_\mu$ are of
order higher to the first, utilizing
the cylindricity condition, we get
\begin{equation}
  u^\nu\nabla_\nu u_\mu=eku^5u^\nu
  F_{\mu\nu}=\frac{\sqrt{4G}}{c^2}
  u^5u^\nu F_{\mu\nu} \label{74}
\end{equation}
which compared with the classical
equation
\begin{equation}
  u^\nu\nabla_\nu u_\mu=\frac{q}{mc^2}
  u^\nu F_{\mu\nu} \label{75}
\end{equation}
gives
\begin{equation}
  u^5=u_5=\frac{q}{2m\sqrt{G}}\,.
  \label{76}
\end{equation}
Now being
\begin{equation}
  p_A\equiv mcu_A   \label{77}                             
\end{equation}
we have what we are looking for that is
the relation between the particle charge
and the fifth component of its 5-momentum:
\begin{equation}
  p_5= \frac{qc}{\sqrt{4G}}\,.   \label{78}                             
\end{equation}

Considering that the fifth dimension is
a circumference of length $L$,
because of the periodicity condition
it must result
\begin{equation}
  p_5=\frac{2\pi n}{L}\hbar \qquad
  \qquad  (n\in \textbf{Z}).  \label{80}                               
\end{equation}
Comparing this result with (\ref{78})
one has
\begin{equation}
  L=4\pi\sqrt{G}\frac{\hbar}{ec}
  \approx2.37\,\,10^{-31}cm \qquad\qquad
  q=ne \label{81} 
\end{equation}
Thus we get the quantization of electric charge and an estimate of
the compattification in agreement with the fact that
extra-dimensions are not observed. Such energies to permit to
explore such distances, where extra-dimensions could yeld relevant
effects, are still not available.\\
Another expression for $L$ will be
obtained subsequently in the attempt
to produce the quantum electrodynamics
and we will see that unless a factor
$\sqrt{4\pi}$ it will coincide with
(\ref{81}).

\subsection{Spinors as Matter Fields and
QED Theory}

Now we introduce matter by spinorial fields in eigenstates of the momentum fifth component
and thus with the following dependence by extra-coordinates
\begin{equation}
  \chi(x^A)=e^{\frac{i}{\hbar}p_5x^5}
  \psi(x^\nu)  \label{79}                                           
\end{equation}
which verify the relations (\ref{c2}) (\ref{f2}) and therefore lead to the conservation
of the charge associated to the group U(1); we define Dirac matrices in the
pentadimensional curved space by their tetradic projections
\begin{equation}
   \gamma^A\equiv V_{(A)}^{\phantom{(A)}A}
   \gamma^{(A)}\qquad \gamma_A\equiv
   V^{(A)}_{\phantom{(A)}A}\gamma_{(A)}\label{106}
\end{equation}
for which we assume the following form
\begin{equation}
\gamma^{(\mu)}=\gamma^{\mu}\qquad\gamma^{(5)}=\gamma_{5}=i\gamma^{0}\gamma^{1}\gamma^{2}\gamma^{3}
\end{equation}
and in this way they satisfy the Dirac algebra
\begin{equation}
[\gamma^{(A)};\gamma^{(B)}]=2I\eta^{(A)(B)}.
\end{equation}

For spinorial connections we assume the standard form for the
four-dimensional ones and (\ref{spconn}) for the
extra-dimensional
\begin{eqnarray}\left\{\begin{array}{cc}
\Gamma_{(\mu)}=^4\Gamma_{(\mu)}\equiv- \frac{1}{4}\gamma^{(\rho)}
\gamma^{(\sigma)}R_{(\sigma)(\rho)(\mu)} \qquad\qquad
\mathrm{greek\,\,indexes \,\,go\,\,from\,\,1\,\,to\,\,4}\qquad
{}\\{} \,\,\Gamma_{(5)}=M\textbf{I}\qquad\qquad
\qquad\qquad\qquad\qquad \mathrm{with\,\,M\,\,constant\,\,and\,\,
\textbf{I}\,\,identity\,\,matrix}. \label{114}
\end{array}\right.\end{eqnarray}

However this choice doesn't affect the four-dimensional theory, in
fact while we do not have
\begin{equation}
   D_A\gamma_B=0  \label{115}                                         
\end{equation}
the Dirac algebra is still valid in the quadridimensional
space-time
\begin{equation}
   ^4D_\mu{}^4\!\gamma_\nu=0;  \label{116}
\end{equation}
moreover the Lagrangian density with this form for the spinorial
connections
\begin{equation}
 \Lambda=-\frac{i\hbar c}{2}\bar{\chi}
   \gamma^{(A)}D_{(A)}\chi+\frac{i\hbar c}{2}
   (D_{(A)}\bar{\chi})\gamma^{(A)}\chi+
   imc^2\bar{\chi}\chi
\end{equation}
 is still invariant under the most general
transformation of coordinate given by (\ref{b1}).\\

By carrying out the dimensional reduction of the Dirac Lagrangian
density one obtains
\begin{eqnarray}
    ^5\Lambda=-\frac{i\hbar c}{2}
    \bar{\chi}\gamma^\mu{}^4\!D_\mu\chi-
    \frac{i\hbar c}{2}({}^4\!D_\mu\bar{\chi})
    \gamma^\mu\chi-\frac{2\pi ek\hbar c}{L}
    A_\mu\bar{\chi}\gamma^\mu\chi+\hbar c(iM+
    \frac{2\pi}{\Phi L})\bar{\chi}\gamma^{(5)}\chi+
    imc^2\bar{\chi}\chi.{}\nonumber\\{} \label{118}
\end{eqnarray}
and imposing
\begin{equation}
   M\equiv\frac{i2\pi}{\Phi L}   \label{119}
\end{equation}
the penultimate term in (\ref{118}) disappears, thus the action is
\begin{equation}
   S=\frac{1}{c}\int\sqrt{-j}\,{}^5\!
   \Lambda\,d^4x\,dx^5 \label{120}
\end{equation}
from which by integrating on $x^5$, using
the cylindricity condition and the
expression of $\chi$ and $\bar{\chi}$, we
get
\begin{eqnarray}
    S=\frac{1}{c}\int\sqrt{-g}\Phi\big[
    -\frac{i\hbar c}{2}\bar{\psi}\gamma^\mu
    D_\mu\psi+\frac{i\hbar c}{2}(D_\mu\bar{\psi})
    \gamma^\mu\psi-\frac{2\pi ek\hbar c}{L}
    A_\mu\bar{\psi}\gamma^\mu\psi+
    imc^2\bar{\psi}\psi\big]\,d^4x\nonumber\\ \label{121}
\end{eqnarray}
where now all the quantities are four-dimensional.\\
Therefore using (\ref{36}) the total action in presence of matter
is
\begin{eqnarray}
    S=\frac{1}{c}\int\sqrt{-g}\bigg[
    -\frac{c^4\Phi}{16\pi G}\widehat{R}-
    \frac{c^4}{8\pi G}\nabla_\mu\partial^\mu\Phi
    -\frac{1}{16\pi}\frac{e^2k^2\Phi^3c^4}
    {4G}F_{\mu\nu}F^{\mu\nu}+
{}\nonumber\\{}
-\frac{i\Phi\hbar c}{2}\bar{\psi}
\gamma^\mu D_\mu\psi+\frac{i\Phi\hbar c}{2}
(D_\mu\bar{\psi})\gamma^\mu\psi-
\frac{2\pi\Phi ek\hbar c}{L}A_\mu\bar{\psi}
\gamma^\mu\psi+
imc^2\Phi\bar{\psi}\psi\bigg]\,d^4x.
{}\nonumber\\{} \label{122}
\end{eqnarray}
We want to stress that what produces the right term is the phase
dependence on the fifth coordinate of the spinor field whereas
what avoid the generation of the term with the electromagnetic
tensor is the particular choice of the spinorial connections.\\

Now we will consider the condition $\Phi=1$ to compare the results
we have obtained with the well-known theory; in this case the
(\ref{122}) becomes
\begin{eqnarray}
    S=\frac{1}{c}\int\sqrt{-g}\bigg[
    -\frac{c^4}{16\pi G}\widehat{R}
    -\frac{1}{16\pi}\frac{e^2k^2c^4}
    {4G}F_{\mu\nu}F^{\mu\nu}-\frac{i\hbar c}{2}\bar{\psi}
\gamma^\mu D_\mu\psi {}\nonumber\\{} +\frac{i\hbar c}{2}
(D_\mu\bar{\psi})\gamma^\mu\psi- \frac{2\pi ek\hbar
c}{L}A_\mu\bar{\psi} \gamma^\mu\psi+
imc^2\bar{\psi}\psi\bigg]\,d^4x\label{122.a}
\end{eqnarray}
where it is clear that we get the ordinary action with the terms
$\frac{1}{16\pi}F_{\mu\nu}F^{\mu\nu}$ and
$eA_\mu\bar{\psi}\gamma^\mu\psi$ by imposing
\begin{eqnarray}\left\{\begin{array}{cc}
\frac{e^2k^2c^4}{4G}=1
{}\\{}
{}\\{}
\frac{2\pi ek\hbar c}{L}=e       \label{123}
\end{array}\right.\end{eqnarray}
from which we have
\begin{eqnarray}\left\{\begin{array}{cc}
k=\frac{\sqrt{4G}}{ec^2}\qquad\qquad\qquad
\qquad\quad\,\,\,
{}\\{}
{}\\{}
L=2\pi\sqrt{4G}\frac{\hbar}{ec}=
4.75\,10^{-31}cm.\label{124}
\end{array}\right.\end{eqnarray}
in agreement with the previous
evaluation.\\

\subsection{$\Phi$ as a Klein-Fock Field
via a Conformal Factor}

In this section we demonstrate that it is possible to interpret
the scalar field in the metric as a Klein-Fock field; in fact we
can carry out a conformal transformation of the four-dimensional
metric $g_{\alpha\beta}$ in such a way to obtain a new scalar
field $\varphi$ with a Klein-Fock field energy-momentum tensor. We
define the conformal transformation
\begin{eqnarray}\left\{\begin{array}{cc}
g_{\alpha\beta}\equiv B^2
(\varphi(x^\nu))\overline{g}_{\alpha\beta}
{}\\{}
g^{\alpha\beta}\equiv\frac{1}
{B^2(\varphi(x^\nu))}
\overline{g}^{\alpha\beta} \label{40}
\end{array}\right.\end{eqnarray}
where $B(\varphi(x^\nu))$ is a function of the scalar field
$\varphi(x^\nu)$, never identically zero and of class at least
$C^2$. Also for the new metric stands the relation
\begin{equation}
\overline{g}_{\alpha\beta}
\overline{g}^{\beta\rho}=
\delta^\rho_\alpha \label{41}
\end{equation}
We redefine $\Phi$
\begin{equation}
 \Phi=A(\varphi)  \label{42}                               
\end{equation}
from which, because of $(\ref{a2})$,
\begin{equation}
j_{55}\equiv A^2(\varphi) \label{43}.                                
\end{equation}
Our aim is to obtain the new lagrangian density and then the new
equations by variational principle. First we have to express the
quantities $R$, $\nabla_\mu\partial^\mu\Phi$ and
$F^{\mu\nu}F_{\mu\nu}$ in the new conformal metric;
four-dimensional Christoffel connections in terms of the old
metric are
\begin{equation}
\Gamma^\rho_{\alpha\beta}\equiv
\frac{1}{2}g^{\rho\sigma}(\partial_\beta
g_{\alpha\rho}+\partial_\alpha  g_{\rho\beta}
-\partial_\rho g_{\alpha \beta}) \label{44}
\end{equation}
from which, by the (\ref{40}), we have
\begin{equation}
\Gamma^\rho_{\alpha\beta}=
\overline{\Gamma}^\rho_{\alpha\beta}+
\frac{B^{'}}{B}(\delta^\rho_\alpha
\partial_\beta\varphi+
\delta^\rho_\beta\partial_\alpha\varphi-
\overline{g}_{\alpha\beta}
\overline{g}^{\rho\sigma}\partial_\sigma\varphi)
\label{45}
\end{equation}
where
\begin{equation}
B^{'}\partial_\beta\varphi=
\partial_\beta B \qquad \mathrm{and}
\qquad B^{'}\equiv\frac{dB}{d\varphi},
\label{46}                                                             
\end{equation}
thus we get
\begin{equation}
\nabla_\beta\partial_\alpha A=
A^{'}\overline{\nabla}_\beta\partial_\alpha
\varphi+\bigg(A^{''}-\frac{2A^{'}B^{'}}{B}\bigg)
(\partial_\alpha\varphi)(\partial_\beta\varphi)
+\frac{A^{'}B^{'}}{B}
\overline{g}_{\alpha\beta}
\overline{g}^{\rho\sigma}(\partial_\sigma\varphi)
(\partial_\rho\varphi) \label{47}
\end{equation}
and as a consequence
\begin{equation}
\nabla_\mu\partial^\mu A=
\frac{A^{'}}{B^2}\overline{\nabla}_\mu
\partial^\mu\varphi+
\bigg(\frac{A^{''}}{B^2}+
\frac{2A^{'}B^{'}}{B^3}\bigg)
\overline{g}^{\mu\nu}
(\partial_\mu\varphi)
(\partial_\nu\varphi) \label{48}
\end{equation}
where the crossed covariant derivatives
are expressed as a function of crossed
Christoffel coefficients.\\
Moreover by the relation which expresses the Ricci's tensor as a
function of Christoffel symbols we get, by a calculation
rather laborious,
\begin{eqnarray}
R_{\alpha\beta}=\overline{R}_{\alpha\beta}
+2\overline{\nabla}_\alpha
\bigg(\frac{B^{'}}{B}
\partial_\beta\varphi\bigg)+
\overline{g}_{\alpha\beta}
\overline{g}^{\rho\sigma}
\overline{\nabla}_\sigma
\bigg(\frac{B^{'}}{B}
\partial_\rho\varphi\bigg)-
2\bigg(\frac{B^{'}}{B}\bigg)^2
(\partial_\alpha\varphi)
(\partial_\beta\varphi)+
{}\nonumber\\{}
+2\bigg(\frac{B^{'}}{B}\bigg)^2
\overline{g}_{\alpha\beta}
\overline{g}^{\rho\sigma}
(\partial_\rho\varphi)
(\partial_\sigma\varphi)\qquad\qquad
\qquad\qquad\qquad\qquad\qquad
\qquad\quad\,\, \label{49}
\end{eqnarray}
from which
\begin{eqnarray}
R=\frac{1}{B^2}
\overline{R}+
\frac{6}{B^2}\overline{g}^{\alpha\beta}
\overline{\nabla}_\alpha
\bigg(\frac{B^{'}}{B}\partial_\beta\varphi\bigg)
+6\bigg(\frac{B^{'}}{B^2}\bigg)^2
\overline{g}^{\alpha\beta}
(\partial_\alpha\varphi)
(\partial_\beta\varphi)=
{}\nonumber\\{}
=\frac{1}{B^2}\overline{R}
+\frac{6B^{'}}{B^3}
\overline{g}^{\alpha\beta}
\overline{\nabla}_\alpha
\partial_\beta\varphi+
\frac{6B^{''}}{B^3}\overline{g}^{\alpha\beta}
(\partial_\alpha\varphi)
(\partial_\beta\varphi)\qquad\qquad
\quad\,\,\, \label{50}
\end{eqnarray}
Finally from the definition of $F_{\mu\nu}$ and $F^{\mu\nu}$
\begin{equation}
F_{\mu\nu}\equiv\nabla_\nu A_\mu-
\nabla_\mu A_\nu=\partial_\nu A_\mu
-\partial_\mu A_\nu \qquad \mathrm{e}
\qquad F^{\mu\nu}=g^{\mu\rho}g^{\nu\sigma}
F_{\rho\sigma}   \label{51}
\end{equation}
it is an immediate consequence
\begin{eqnarray}\left\{\begin{array}{ccc}
F_{\mu\nu}=\overline{F}_{\mu\nu}\quad
{}\\{}F^\mu_{\phantom{\mu}\nu}=
\frac{1}{B^2}\overline{F}^\mu_{\phantom{\mu}\nu}
{}\\{}
F^{\mu\nu}=\frac{1}{B^4}
\overline{F}^{\mu\nu}   \label{52}                   
\end{array}\right..\end{eqnarray}
Therefore the lagrangian density in terms of the new
four-dimensional metric is
\begin{eqnarray}
    \Lambda=-\frac{c^4A}{16\pi GB^2}\bigg
[\overline{R}+2\bigg(\frac{A^{'}}{A}
+3\frac{B^{'}}{B}\bigg)
\overline{g}^{\alpha\beta}
\overline{\nabla}_\alpha\partial_\beta\varphi
+\bigg(2\frac{A^{''}}{A}+4\frac{A^{'}B^{'}}{AB}
+6\frac{B^{''}}{B}\bigg)\overline{g}^{\alpha\beta}
(\partial_\alpha\varphi)(\partial_\beta\varphi)+
{}\nonumber\\{}
+\frac{e^2k^2A^2}{4B^2}\overline{F}_{\mu\nu}
\overline{F}^{\mu\nu}\bigg]\qquad\qquad\qquad
\qquad\qquad\qquad
\qquad\qquad\qquad\qquad\qquad\qquad\qquad
\qquad\quad\,\,\,      \label{53}
\end{eqnarray}                                             
where the two degrees of freedom introduced by the functions
$B(\varphi)$ and $A(\varphi)$ can be utilized by imposing
\begin{equation}
\frac{A^{'}}{A}=-2\frac{B^{'}}{B}\qquad
\Rightarrow\qquad\left\{\begin{array}{cc}
\frac{A^{''}}{A}=6\bigg(\frac{B^{'}}{B}\bigg)^2
-2\frac{B^{''}}{B}{}\\{}A=\frac{\lambda}{B^2}
\qquad\qquad\qquad\end{array}\right.  \label{54}
\end{equation}                                                   
with $\lambda$ constant, which leads to
\begin{eqnarray}
   \Lambda=-\frac{c^4\lambda}{16\pi GB^4}
\bigg[\overline{R}+ 2\bigg(\frac{B^{'}}{B}\bigg)
\overline{g}^{\alpha\beta}
\overline{\nabla}_\alpha\partial_\beta\varphi+
\bigg(4\Big(\frac{B^{'}}{B}\Big)^2+ 2\frac{B^{''}}{B}\bigg)
\overline{g}^{\alpha\beta} (\partial_\alpha\varphi)
(\partial_\beta\varphi)+ {}\nonumber\\{}
+\frac{e^2k^2\lambda^2}{4B^{6}}
\overline{F}_{\mu\nu}\overline{F}^{\mu\nu}\bigg]
\qquad\qquad\qquad\qquad\qquad\quad
\qquad\qquad\qquad\qquad\qquad\qquad\quad\,. \label{55}
\end{eqnarray}
We observe that from (\ref{40}) we have
\begin{equation}
   B^4\sqrt{-\overline{g}}=\sqrt{-g}.
    \label{57}
\end{equation}
and thus the action is
\begin{equation}
 S=\frac{B^4}{c}\int\sqrt{-\overline{g}}
   \Lambda\,d\Omega; \label{56}
\end{equation}
now by an integration by parts and by using the relation
$\overline{\nabla}_\mu\partial^\mu\varphi=
\frac{1}{\sqrt{-\overline{g}}}
\partial_\mu(\sqrt{-\overline{g}}
\partial^\mu\varphi)$ we get
\begin{equation}
\int\sqrt{-\overline{g}}\Big
[2\bigg(\frac{B^{'}}{B}\bigg)
\overline{g}^{\alpha\beta}
\overline{\nabla}_\alpha\partial_\beta\varphi\Big]
\,d\Omega
=\int\sqrt{-\overline{g}}
\Big[2\bigg(\Big(\frac{B^{'}}{B}\Big)^2
-\frac{B^{''}}{B}\bigg)
\overline{g}^{\alpha\beta}(\partial_\alpha\varphi)
(\partial_\beta\varphi)\Big]\,d\Omega  \label{58}
\end{equation}
and our action becomes
\begin{equation}
S=-\frac{c^3\lambda}{16\pi G}\int
\sqrt{-\overline{g}}\Big
[\overline{R}+6\Big(\frac{B^{'}}{B}\Big)^2
\overline{g}^{\alpha\beta}(\partial_\alpha\varphi)
(\partial_\beta\varphi)+\frac{e^2k^2\lambda^2}{4B^{6}}
\overline{F}_{\mu\nu}\overline{F}^{\mu\nu}\Big]
\,d\Omega.\label{59}
\end{equation}
At this point to get fields equations we consider arbitrary
variations of the fields $\overline{g}^{\alpha\beta}$, $A_\mu$ and
the scalar field $\varphi$
\begin{eqnarray}
    \delta S=-\frac{c^3\lambda}{16\pi G}
    \int\sqrt{-\overline{g}}\bigg\{\Big[
    \overline{G}_{\alpha\beta}+
    6\Big(\frac{B^{'}}{B}\Big)^2
    \big((\partial_\alpha\varphi)
    (\partial_\beta\varphi)-
    \frac{1}{2}\overline{g}_{\alpha\beta}
    \overline{g}^{\rho\sigma}
    (\partial_\rho\varphi)(\partial_\sigma\varphi)
    \big)+{}\nonumber\\{}+\frac{\lambda^2e^2k^2}{2B^6}
    \big(\overline{F}_{\alpha\sigma}
    \overline{F}_\beta^{\phantom{\beta}\sigma}
    -\frac{\overline{g}_{\alpha\beta}}{4}
    \overline{F}_{\mu\nu}\overline{F}^{\mu\nu}\big)
    \Big]\delta \overline{g}^{\alpha\beta}-\Big[12
    \bigg(\frac{B^{'}B^{''}}{B^2}-
    \Big(\frac{B^{'}}{B}\Big)^3\bigg)
    \overline{g}^{\rho\sigma}(\partial_\rho\varphi)
    (\partial_\sigma\varphi)+
    {}\nonumber\\{}
    +12\bigg(\frac{B^{'}}{B}\bigg)^2
    \overline{\nabla}_\mu\partial^\mu\varphi+
    \frac{3}{2}\frac{\lambda^2e^2k^2B^{'}}{B^7}
    \overline{F}_{\mu\nu}\overline{F}^{\mu\nu}\Big]
    \delta\varphi+\Big[\frac{\lambda^2e^2k^2}{4}
    \overline{\nabla}_\nu(\frac{1}{B^6}
    \overline{F}^{\nu\mu})\Big]\delta A_\mu
    \bigg\}\,d\Omega \label{60}
\end{eqnarray}
and then we have
\begin{eqnarray}\left\{\begin{array}{ccc}
\overline{G}_{\alpha\beta}+6\Big(\frac{B^{'}}{B}
\Big)^2\big((\partial_\alpha\varphi)
(\partial_\beta\varphi)-\frac{1}{2}
\overline{g}_{\alpha\beta}
\overline{g}^{\rho\sigma}(\partial_\rho\varphi)
(\partial_\sigma\varphi)\big)+
\frac{\lambda^2e^2k^2}{2B^6}
\big(\overline{F}_{\alpha\sigma}
\overline{F}_\beta^{\phantom{\beta}\sigma}
-\frac{\overline{g}_{\alpha\beta}}{4}
\overline{F}_{\mu\nu}\overline{F}^{\mu\nu}\big)=0
{}\\{}
\bigg(\frac{B^{''}}{B}-
\Big(\frac{B^{'}}{B}\Big)^2\bigg)
\overline{g}^{\rho\sigma}(\partial_\rho\varphi)
(\partial_\sigma\varphi)+
\bigg(\frac{B^{'}}{B}\bigg)
\overline{\nabla}_\mu\partial^\mu\varphi
+\frac{1}{8}\frac{\lambda^2e^2k^2}{B^6}
\overline{F}_{\mu\nu}\overline{F}^{\mu\nu}=
0\qquad\qquad\qquad\qquad\,\,\,
{}\\{}
\overline{\nabla}_\nu(\frac{1}{B^6}
\overline{F}^{\nu\mu})=0\qquad\qquad
\qquad\qquad\qquad\qquad
\qquad\qquad\qquad\qquad\qquad\qquad
\qquad\qquad\qquad\,\,
\label{61}
\end{array}\right.\end{eqnarray}
To get equations of the same kind as Einstein ones in presence of
matter we use the degree of freedom given by the function B, in
particular we impose
\begin{equation}
   6\Big(\frac{B^{'}}{B}\Big)^2=
   \frac{\lambda^2}{B^6}  \label{63}
\end{equation}
which allows the solution
\begin{equation}
   B=D\sqrt[3]{\varphi}   \label{64}
\end{equation}
and from (\ref{61}), using the relation (\ref{73}), we have
\begin{equation}
\overline{G}_{\alpha\beta}=\frac{8\pi G}{c^4}
\frac{2}{3\varphi^2}\Big\{-\frac{c^4}{8\pi G}
\frac{1}{2}\Big[(\partial_\alpha\varphi)
(\partial_\beta\varphi)-\frac{1}{2}
\overline{g}_{\alpha\beta}
\overline{g}^{\rho\sigma}(\partial_\rho\varphi)
(\partial_\sigma\varphi)\Big]-
\frac{1}{4\pi}\big(\overline{F}_{\alpha\sigma}
\overline{F}_\beta^{\phantom{\beta}\sigma}-
\frac{\overline{g}_{\alpha\beta}}{4}
\overline{F}_{\mu\nu}\overline{F}^{\mu\nu}
\big)\Big\}, \label{66}
\end{equation}
We can interpret the last equation assuming that G, Newton's
gravitational constant, in some way varies in the space and in the
time on large scale according to
\begin{eqnarray}
    G^*=\frac{2G}{3\varphi^2}
\end{eqnarray}
where $\varphi$ is a cosmological Klein-Fock field.

\subsection{The Gravity-Matter action and
the Conformal Factor}

In this section we present the problems that arise by getting the
conformal transformation on the four-dimensional metric in
presence of matter and we show how they can be overcome, thus
obtaining a theory with gravity, a Klein-Fock field with a
geometric source and spinorial matter fields.\\
From (\ref{40}) we get the following relations between tetradic
vectors of the old metric and the new ones
\begin{equation}
   V^{(\alpha)}_{\phantom{(\alpha)}\,\,\alpha}=
   B\overline{V}^{(\alpha)}_{\phantom{
   (\alpha)}\,\,\alpha}             \label{126}
\end{equation}
\begin{equation}
   V_{(\alpha)}^{\phantom{(\alpha)}\,\,\alpha}=
   \frac{1}{B}\overline{V}_{(\alpha)}^
   {\phantom{(\alpha)}\,\,\alpha}\label{127}
\end{equation}
and we thus obtain
\begin{eqnarray}\left\{\begin{array}{cc}
\overline{V}^{(5)}_{\phantom{(5)}\,\,5}=
\frac{\lambda}{B^2}\qquad\,
{}\\{}
\overline{V}^{(5)}_{\phantom{(5)}\,\,\mu}=
ekA_\mu\frac{\lambda}{B^2}
{}\\{}
\overline{V}_{(5)}^{\phantom{(5)}\,\,5}=
\frac{B^2}{\lambda}\qquad.\label{128}
\end{array}\right.\end{eqnarray}
while the components which were zero previously
remain zero.\\
To get the new spacetemporal Dirac's matrices we start from the
tetradic ones that are not modified by the conformal
transformation, so we have
\begin{equation}
   \overline{\gamma}^{\,\mu}\equiv\,
   \overline{V}_{(\rho)}^{\phantom{(\rho)}
   \,\,\mu}\gamma^{(\rho)}=B\gamma^\mu
   \label{129}
\end{equation}
and
\begin{equation}
   ^4\overline{\gamma}_\mu=
   \overline{V}^{(\rho)}_{\phantom{(\rho)}\,\,\mu}
   \gamma_{(\rho)}=\frac{1}{B}\,{}^4\gamma_\mu
   \label{130}
\end{equation}
from which we obtain for spinorial connections
\begin{equation}
   \Gamma_\mu=-\frac{1}{4}\gamma^\rho
    \nabla_\mu\gamma_\rho=\overline{\Gamma}_\mu+
   \frac{1}{4}
   \frac{B^{'}}{B}[\overline{\gamma}^
   {\,\rho},\overline{\gamma}_\mu]
   \partial_\rho\varphi.       \label{132}                          
\end{equation}
Now, we can rewrite the lagrangian density of the action
(\ref{121}) as
\begin{eqnarray}
    \Lambda=-\frac{i\hbar c}{2B}\bar{\psi}
    \overline{\gamma}^{\,\mu}
    \overline{D}_\mu\psi+\frac{i\hbar c}{2B}
    (\overline{D}_\mu\bar{\psi})
    \overline{\gamma}^{\,\mu}\psi+
    \frac{i\hbar c}{4}\frac{B^{'}}{B^2}\bar{\psi}
    \{\overline{\gamma}^{\,\mu},
    \overline{\Sigma}^{\,\rho}_{\phantom{\rho}\mu}\}
    \psi(\partial_\rho\varphi)+
{}\nonumber\\{}
-\frac{e}{B}A_\mu\bar{\psi}
\overline{\gamma}^{\,\mu}\psi+imc^2\bar{\psi}
\psi\qquad\qquad\qquad\qquad\qquad\qquad
\qquad\qquad\,\,\,\,\,\label{133}
\end{eqnarray}
where
\begin{equation}
   \overline{\Sigma}^{\,\rho}_{\phantom{\rho}\mu}
   \equiv\frac{1}{2}
   [\overline{\gamma}^{\,\rho},
   \overline{\gamma}_\mu]
      \label{134}
\end{equation}
but, by using the anticommutation rules for the new Dirac's
matrices, it is immediate to show that
\begin{equation}
   \{\overline{\gamma}^{\,\mu},
   \overline{\Sigma}^{\,\rho}_
   {\phantom{\rho}\mu}\}=0.\label{135}
\end{equation}
So our action is
\begin{eqnarray}
    S=\frac{1}{c}\int\sqrt{-\overline{g}}
    \,\Big\{-\frac{i\hbar\lambda cB}{2}\bar{\psi}
    \overline{\gamma}^{\,\mu}
    \overline{D}_\mu\psi+
    \frac{i\hbar\lambda cB}{2}
    (\overline{D}_\mu\bar{\psi})
\overline{\gamma}^{\,\mu}\psi-eB\lambda
A_\mu\bar{\psi}\overline{\gamma}^{\,\mu}\psi+
imc^2B^2\lambda\bar{\psi}\psi\Big\}\,
d^4x{}\nonumber\\{}            \label{136}
\end{eqnarray}
and to get the field equations, we can proceed by varying the
(\ref{136}) and we obtain in such a way for the spinorial field
\begin{eqnarray}\left\{\begin{array}{cc}
i\hbar c\lambda B(\overline{D}_\mu\bar{\psi})
\overline{\gamma}^{\,\mu}+\frac{i\hbar c\lambda}{2}
B^{'}(\partial_\mu\varphi)
\bar{\psi}\overline{\gamma}^{\,\mu}-eB\lambda
A_\mu\bar{\psi}\overline{\gamma}^{\,\mu}+
imc^2B^2\lambda\bar{\psi}=0 {}\\{} {}\\{} i\hbar c\lambda B
\overline{\gamma}^{\,\mu}\overline{D}_\mu\psi +\frac{i\hbar
c\lambda}{2}B^{'}
(\partial_\mu\varphi)\overline{\gamma}^{\,\mu}\psi +eB\lambda
A_\mu\overline{\gamma}^{\,\mu}\psi-
imc^2B^2\lambda\psi=0;\,\,\label{137}
\end{array}\right.\end{eqnarray}
these equations bring to the non-conservation of the charge, in
fact it is easy to get from (\ref{137})
\begin{equation}
   \overline{D}_\mu\overline{j}^{\,\mu}=
   -\frac{B^{'}}{B}(\partial_\mu\varphi)
   \overline{j}^{\,\mu}.\label{138}
\end{equation}
Such an inconvenience can be overcome by a conformal
transformation on the spinorial field
\begin{equation}
   \psi=F(\varphi)\psi^* \qquad\qquad
   \bar{\psi}=F(\varphi)\bar{\psi}^*
   \label{139}
\end{equation}
that brings to a new action given by (\ref{136}) in which every
term is multiplied by a factor $F^2(\varphi)$; in particular the
terms  $\,\,\, \frac{i\hbar c\lambda}{2}B^{'}
(\partial_\mu\varphi) \overline{\gamma}^{\,\mu}\psi\,\,\,$ and
$\,\,\,\frac{i\hbar c\lambda}{2} B^{'}(\partial_\mu\varphi)
\bar{\psi}\overline{\gamma}^{\,\mu}\,\,\,$ which produce the
(\ref{138}), become $\,\,\,\frac{i\hbar c\lambda}{2}
(BF^2)^{'}(\partial_\mu\varphi)
\overline{\gamma}^{\,\mu}\psi\,\,\,$ and $\,\,\,\frac{i\hbar
c\lambda}{2} (BF^2)^{'}(\partial_\mu\varphi)
\bar{\psi}\overline{\gamma}^{\,\mu}\,\,\,$ and they disappear by
imposing
\begin{equation}
   (BF^2)^{'}=0       \label{140}
\end{equation}
which leads to
\begin{eqnarray}
   F=\frac{\Lambda}{\sqrt{B}}\label{141}
\end{eqnarray}
where $\Lambda$ is a constant.\\
With such a condition, the spinor equations
become
\begin{eqnarray}\left\{\begin{array}{cc}
i\hbar c\lambda \overline{\gamma}^{\,\mu}
\overline{D}_\mu\psi+e\lambda
A_\mu\overline{\gamma}^{\,\mu}\psi-imc^2B
\lambda\psi=0\,\,\,\,\,
{}\\{}
i\hbar c\lambda (\overline{D}_\mu\bar{\psi})
\overline{\gamma}^{\,\mu}-e\lambda
A_\mu\bar{\psi}\overline{\gamma}^{\,\mu}+
imc^2B\lambda\bar{\psi}=0\label{142}
\end{array}\right.\end{eqnarray}
while the other field equations with $k=2\sqrt{G}/ec^2$ are
\begin{eqnarray}\left\{\begin{array}{cc}
\overline{G}_{\alpha\beta}=
\frac{8\pi G}{c^4}\bigg[6\Big(\frac{B^{'}}{B}\Big)^2
T^{\phi}_{\alpha\beta}+\frac{\lambda^2}{B^6}
T^{EM}_{\alpha\beta}+\lambda \Lambda
T^{(Spin+inter)}_{\alpha\beta}\bigg]\qquad
\qquad\qquad\qquad\,\qquad
{}\\{}
\bigg(\frac{B^{''}}{B}-
\Big(\frac{B^{'}}{B}\Big)^2\bigg)
\overline{g}^{\rho\sigma}\partial_\rho\varphi
\partial_\sigma\varphi+
\frac{B^{'}}{B}\overline{\nabla}_\rho
\partial^\rho\varphi+\frac{G}{2c^4}
\frac{\lambda^2}{B^6}\overline{F}_{\mu\nu}
\overline{F}^{\,\mu\nu}-
i\frac{4\pi GmB\lambda}{3c^2}\bar{\psi}\psi=
0\qquad
{}\\{}
\overline{\nabla}_\nu
\frac{\lambda^2\overline{F}^{\,\nu\mu}}
{B^6}=\lambda \Lambda\frac{4\pi}{c}
\overline{j}^\mu \qquad\qquad\qquad
\qquad\qquad\qquad\qquad
\qquad\qquad\qquad\qquad\,\,\qquad\label{143}
\end{array}\right.\end{eqnarray}
where
\begin{eqnarray}\left\{\begin{array}{cc}
T^{\varphi}_{\alpha\beta}=-\frac{c^4}{8\pi G}
\Big(\partial_\alpha\varphi
\partial_\beta\varphi-
\frac{\overline{g}_{\alpha\beta}}{2}
\overline{g}^{\,\rho\sigma}\partial_\rho
\varphi\partial_\sigma\varphi\Big)\qquad
\qquad\qquad\qquad\quad
{}\\{}
T^{EM}_{\alpha\beta}=-\frac{1}{4\pi}
\Big(\overline{F}_{\alpha\sigma}
\overline{F}_{\beta}^{\phantom{\beta}\sigma}
-\frac{\overline{g}_{\alpha\beta}}{4}
\overline{F}_{\mu\nu}\overline{F}^{\,\mu\nu}\Big)
\qquad\qquad\qquad\qquad\quad\,\,\,\,\,\,\,
{}\\{}
T^{(Spin+inter)}_{\alpha\beta}=
[\frac{i\hbar c}{2}\bar{\psi}
\overline{\gamma}_{(\alpha}
\overline{D}_{\beta)}\psi-\frac{i\hbar c}{2}
(\overline{D}_{(\alpha}\bar{\psi})
\overline{\gamma}_{\beta)}\psi+
eA_{(\alpha}\bar{\psi}
\overline{\gamma}_{\beta)}\psi].\label{144}
\end{array}\right.\end{eqnarray}
We suppose the validity of the (\ref{63})
and then of the (\ref{64}) from which we
have
\begin{equation}
   D=\Big(\frac{3\lambda^2}{2}\Big)^{\frac{1}{6}}
   \label{145}
\end{equation}
and now it is easy to see that setting
\begin{equation}
   \lambda=1 \qquad\qquad\qquad \Lambda=1
\end{equation}
we obtain:
\begin{itemize}\item Einstein's equations in which $G$
depends on the space-time coordinates and there is a conformal
factor in front of the energy-momentum tensor of the matter;
\item Maxwell's equations in which the electromagnetic tensor is
coupled to the scalar field's gradient;\item the equation describing 
the dynamics of the field $\varphi$, which does not resemble any known motion equation.\end{itemize}

\subsection{Cosmological Implementation and the
Dirac idea}

In the last section we will discuss about
a possible spacetemporal dependence of some
physical quantities such as the electric
charge or Newton's gravitational constant.\\
In order to do this we consider both the Dirac's work about his
Large Number Hypothesis \cite{D82} and a Chodos and Detweiler work
\cite{CD80}. In the latter they assume a Kasner metric to describe
their five-dimensional universe and show a suitable scenario in
which the compactification process is due exclusively to the
temporal
evolution of the universe.\\
We will assume the Kasner metric as well to describe our
5-dimensional universe, identifying so the Klein-Fock scalar field
as a time depending function. This will allow us to deduce a
temporal dependence of fundamental constants by the analysis of
four-dimensional Einstein's equations and Dirac's equations for
the spinorial field and its
adjoint.\\
As Chodos and Detweiler do, we assume the following
Kasner metric:
\begin{equation}
  ds^2=-dt^2+\sum^{d}_{i=1}(t/t_0)^{2p_{i}}(dx^i)^2
\label{146}
\end{equation}
with the condition on $p_i$
\begin{equation}
  \sum^{d}_{i=1}p_i=\sum^{d}_{i=1}p_i^2=1.
\label{147}
\end{equation}
In order to guarantee the spatial isotropy of
the universe we can chose
\begin{equation}
  p_1=p_2=p_3=\frac{1}{2}\qquad p_5=-\frac{1}{2},
\label{148}
\end{equation}
obtaining so
\begin{equation}
  ds^2=-dt^2+(t/t_0)[(dx^1)^2+(dx^2)^2+(dx^3)^2]
  +(t_0/t)(dx^5)^2.
\label{149}
\end{equation}
It is clear that for $t=t_0$ the universe has four spatial
dimensions of the same length. Instead for $t<<t_0$ it has
essentially only one spacial dimension whereas for $t>>t_0$ it
assumes the current configuration with three spatial dimensions
and a fifth one of length
$L^{'}=(t_0/t)^{1/2}L$.\\
By rewriting the line element (\ref{149}) pointing out the
conformal transformation we have
\begin{equation}
  ds^2=B^2(\varphi)\big[-\frac{dt^2}{B^2(\varphi)}+
  \frac{t}{t_0}\frac{d\vec{l}^2}{B^2(\varphi)}\big]+
  \frac{t_0}{t}(dx^5)^2
\label{150}
\end{equation}
where there aren't the electromagnetic terms because
we consider them as a perturbation.\\
Thanks to expression (\ref{57}) and to the relation
between the old and the new scalar field we can write
\begin{equation}
   \frac{\lambda^2}{B^4(\varphi)}=\frac{t_0}{t}
\label{151}
\end{equation}
from which, knowing that $B^4\propto
\varphi^{4/3}$, we can affirm that
\begin{equation}
  \varphi^{-4/3}\propto(t_0/t)\qquad\Rightarrow
  \qquad  \varphi\propto(t_0/t)^{-3/4}.
\label{152}
\end{equation}
Now we consider the Einstein's equations
and Dirac's equations in the conformal reference
frame.
From the former we obtain
\begin{equation}
   G\propto\varphi^{-2}\qquad\Rightarrow\qquad
   G\propto(t_0/t)^{3/2}
\label{153}
\end{equation}
while from the latter, considering the speed of
light as a constant, we can argue the constancy of
the electric charge and the dependence on time of
masses in accordance with the law
\begin{equation}
m\propto B(\varphi)\propto(t_0/t)^{-1/4}.
\label{154}
\end{equation}
Now if we analyze the ratio estimated by Dirac we obtain
\begin{equation}
e^2/m^2G\propto t
\label{155}
\end{equation}
in accordance with his hypothesis.\\
Nevertheless, $t$ is the age of the universe
only for a five-dimensional observer whereas an
observer linked to the conformal reference frame
will measure $\tau$ as age of the universe that
as we will see is proportional to $t^{3/4}$.\\
To show the last statement we reconsider the
(\ref{150}) from which it is clear that the
coefficient of $dt^2$ in the square brackets
is proportional to $1/\sqrt{t}$.\\
Now we define the variable $\tau$ in such a
way that the metric become
\begin{equation}
ds^2=B^2(\varphi)\big[-d\tau^2+
f(\tau)d\vec{l}^2\big]+g(\tau)(dx^5)^2
\label{156}
\end{equation}
where $g$ and $f$ depend on $\tau$.\\
It is easy to show that the condition
\begin{equation}
d\tau^2=\frac{dt^2}{B^2(\varphi)}
\label{157}
\end{equation}
implies
\begin{equation}
\tau\propto t^{3/4}.
\label{158}
\end{equation}
So the ratio between the electric and the
gravitational attraction for a four-dimensional
observer is proportional to $\tau^{4/3}$ not
according with Dirac's hypothesis.\\

\newpage

\section{Geometrization of electro-weak model}

\subsection{Space-time manifold and standard model particles}
Let consider the application to a space-time
manifold $V^{4}\otimes S^{1}\otimes S^{2}$; the
extra-dimensional space is such that it is possible to take one
killing vector on $S^{1}$ and three Killing vectors on $S^{2}$
whose algebra is the same of the group
$SU(2)\otimes U(1)$ that therefore can be geometrized in this manifold in a Kaluza-Klein approach.\\
In this case we introduce three $\alpha$ fields, one describing
$S^{1}$ ($\alpha'$) and two $S^{2}$, that, for simplicity, we assume equal ($\alpha$).\\
From the dimensional reduction of the seven-dimensions
Einstein-Hilbert action we get the ordinary Einstein-Hilbert
action, the Yang-Mills action for the four gauge bosons plus the
terms that describe the dynamics of the fields $\alpha$ (\ref{azsplit}).\\
We introduce matter by spinorial fields; in particular we
take 8-components spinors and this choice fixes univocally the
number of space-time dimensions. In fact, to geometrize
$SU(2)\otimes U(1)$ we need at least of an adjunctive three
dimensional space, like $S^{1}\otimes S^{2}$, but to define Dirac
algebra are necessary n matrix which anticommute among them and
the maximum number of $8\times8$ matrix that anticommute is 7;
thus we must consider a seven dimensions space time manifold.\\
In particular we have the following representation for Dirac
matrices
\begin{eqnarray*}
\gamma^{(0)}=\left(\begin{array}{cc} \gamma^{0} & 0 \\
 0 & \gamma^{0}\end{array}\right)\qquad\gamma^{(1)}=\left(\begin{array}{cc} \gamma^{1} & 0 \\
 0 & \gamma^{1}\end{array}\right)\\\\\gamma^{(2)}=\left(\begin{array}{cc} \gamma^{2} & 0 \\
 0 & \gamma^{2}\end{array}\right)\qquad
 \gamma^{(3)}=\left(\begin{array}{cc} \gamma^{3} & 0 \\
 0 & \gamma^{3}\end{array}\right)\\\\\gamma^{(4)}=\left(\begin{array}{cc} \gamma_{5} & 0 \\
 0 & -\gamma_{5}\end{array}\right)\qquad\gamma^{(5)}=\left(\begin{array}{cc} 0 & \gamma_{5} \\
  \gamma_{5} & 0 \end{array}\right)\\\\\gamma^{(6)}=\left(\begin{array}{cc} 0 & i\gamma_{5} \\
  -i\gamma_{5} & 0 \end{array}\right).
\end{eqnarray*}
Moreover, we can reproduce all the particles of standard
electro-weak model by two spinors for every leptonic family
and for every quark generation
\begin{equation}
\Psi_{lL}=\frac{1}{\sqrt{V^{K}}}e^{-iT_{\bar{i}}\lambda^{\bar{i}}_{(L)\bar{Q}}\Theta^{\bar{Q}}}
\left(\begin{array}{c}e^{-i\lambda^{\bar{0}}_{(L)\bar{Q}}\Theta^{\bar{Q}}}\psi_{\nu_{l}L}\\
e^{-i\lambda^{\bar{0}}_{(L)\bar{Q}}\Theta^{\bar{Q}}}\psi_{lL}\end{array}\right)\qquad
\Psi_{lR}=\frac{1}{\sqrt{V^{K}}}\left(\begin{array}{c}\psi_{\nu_{l}R}\\
e^{-i\lambda^{\bar{0}}_{(lR)\bar{Q}}\Theta^{\bar{Q}}}\psi_{lR}\end{array}\right)
\end{equation}
\begin{equation}
\Psi_{gL}=\frac{1}{\sqrt{V^{K}}}e^{-iT_{\bar{i}}\lambda^{\bar{i}}_{(L)\bar{Q}}\Theta^{\bar{Q}}}
\left(\begin{array}{c}e^{-i\lambda^{\bar{0}}_{(qL)\bar{Q}}\Theta^{\bar{Q}}}u_{gL}\\
e^{-i\lambda^{\bar{0}}_{(qL)\bar{Q}}\Theta^{\bar{Q}}}d_{gL}\end{array}\right)\qquad
\Psi_{qR}=\frac{1}{\sqrt{V^{K}}}\left(\begin{array}{c}e^{-i\lambda^{\bar{0}}_{(uR)\bar{Q}}\Theta^{\bar{Q}}}u_{gR}\\
e^{-i\lambda^{\bar{0}}_{(dR)\bar{Q}}\Theta^{\bar{Q}}}d_{gR}\end{array}\right)
\end{equation}\\
where for $\lambda$ we have
\begin{equation}\label{deflam}
a_{\ldots\bar{Q}}(\lambda^{-1}_{\ldots})^{\bar{P}}_{\bar{Q}}=\frac{1}{V^{K}}\int_{B^{K}}\sqrt{-\gamma}\Big(\xi^{m}_{\bar{Q}}\partial_{m}\Theta^{\bar{P}}\Big)d^{K}y
\end{equation}
with
\begin{eqnarray}
a_{L\bar{0}}=-\frac{1}{2} \qquad
a_{L\bar{1}}=a_{L\bar{2}}=a_{L\bar{3}}=1\\
a_{lR\bar{0}}=-1\\
a_{qL\bar{0}}=+\frac{1}{6} \qquad
a_{qL\bar{1}}=a_{qL\bar{2}}=a_{qL\bar{3}}=1\\
a_{uR\bar{0}}=\frac{2}{3}\qquad a_{dR}=-\frac{1}{3}.
\end{eqnarray}
and $\frac{1}{2}T_{\bar{i}}$ are $SU(2)$ generators in an
eight-dimensional representation and can be taken as
\begin{equation}
T_{\bar{1}}=\left(\begin{array}{cc} 0 & I
\\ I & 0 \end{array}\right)\qquad T_{\bar{2}}=\left(\begin{array}{cc} 0 & -iI \\ iI & 0
\end{array}\right)\qquad T_{\bar{3}}=\left(\begin{array}{cc} I & 0
\\ 0 & -I \end{array}\right)
\end{equation}\\
where I is four-dimensional identity.\\
Now, if we consider translations along extra-dimensions given by
the second of (\ref{b1}), we found that the leptonic fields
transform as
\begin{eqnarray}
\psi'_{L}=\bigg(I-i\frac{1}{2}\delta\omega^{\bar{0}}I
+i\frac{1}{2}\delta\omega^{\bar{i}}T_{\bar{i}}\bigg)\psi_{L}\\
\psi'_{lR}=(I-i\delta\omega^{\bar{0}}I)\psi_{lR}\qquad\qquad\quad\\
\psi'_{\nu_{l}R}=\psi_{\nu_{l}R}\qquad\qquad\qquad\qquad\quad
\end{eqnarray}
which are the correct transformations laws in standard model \cite{B8}; one
can verify that the same result stands for quarks.\\
So, in this model we don't have the chirality problem, which
typically affects theories which deal with electro-weak model
geometrization \cite{W81} \cite{W83}; this because in those
theories they associate gauge transformations to rotations of
multi-dimensional tetrad and look for complex irreducible
representations of O(1; n-1) which are solution of
\begin{equation}
\gamma^{m}\partial_{m}\Psi=0.
\end{equation}
In our case, instead, \emph{gauge transformations are reproduced
by translations and the different transformation proprieties
between left-handed and right-handed states descends not from the
fact that they are complex representations of any group but from a
different
dependence by extra-dimensional coordinates}.\\
\\
Thus the total action is
\begin{eqnarray*}
S=-\frac{1}{c}\int \bigg[\frac{c^{4}}{16\pi
G_{(7)}}{}^7\!R+\sum_{l=1}^{3}\bigg(\frac{i\hbar
c}{2}(D_{\bar{A}}\bar{\Psi}_{lL}\gamma^{\bar{A}}\Psi_{lL}-\bar{\Psi}_{lL}\gamma^{\bar{A}}D_{\bar{A}}\Psi_{lL}+\\+
D_{\bar{A}}\bar{\Psi}_{lR}\gamma^{\bar{A}}\Psi_{lR}-\bar{\Psi}_{lR}\gamma^{\bar{A}}D_{\bar{A}}\Psi_{lR})\bigg)
+\frac{i\hbar
c}{2}\sum_{g=1}^{3}\bigg(D_{A}\bar{\Psi}_{gL}\gamma^{A}\Psi_{gL}
-\\-\bar{\Psi}_{gL}\gamma^{A}D_{A}\Psi_{gL}+
D_{A}\bar{\Psi}_{gR}\gamma^{A}\Psi_{gR}-
\bar{\Psi}_{lR}\gamma^{A}D_{A}\Psi_{gR}\bigg)\bigg]\sqrt{-g}\sqrt{-\gamma}d^{4}x
d^{3}y
\end{eqnarray*}
where the sums run over the three leptonic families and the three
quark generations; we, now, insert the coupling constants by
redefining gauge bosons
\begin{equation}
A^{\bar{0}}_{\mu}\Rightarrow kg'B_{\mu}\qquad
A^{\bar{i}}_{\mu}\Rightarrow kg W^{\bar{i}}_{\mu}\quad (i=1,2,3)
\end{equation}
and in this way, after the dimensional splitting with the suitable
choice for spinorial connections, we have
\begin{eqnarray*}
S=-\frac{1}{c}\int d^{4}x \sqrt{-g}\bigg[\frac{c^{4}}{16\pi
G}R+\frac{c^{4}}{16\pi
G}\bigg(-2g^{\mu\nu}\sum_{n=1}^{3}\frac{\nabla_{\mu}\partial_{\nu}\alpha^{n}}{\alpha^{n}}-g^{\mu\nu}\sum_{n\neq
m=1}^{3}\frac{\partial_{\mu}\alpha^{m}}{\alpha^{m}}\frac{\partial_{\nu}\alpha^{n}}{\alpha^{n}}\bigg)+\\\\+\frac{c^{4}}{16\pi
G}\frac{1}{4}k^{2}g^{2}(\alpha')^{2}\sum_{i=1}^{3}W^{i}_{\mu\nu}W^{i\mu\nu}+\frac{c^{4}}{16\pi
G}\frac{1}{4}k^{2}g'^{2}(\alpha')^{2}B_{\mu\nu}B^{\mu\nu}+\sum_{l=1}^{3}\bigg(\frac{i\hbar
c}{2}(D^{(4)}_{\bar{\mu}}\bar{\psi}^{l}_{L}\gamma^{\bar{\mu}}\psi^{l}_{L}-\\\\-
\bar{\psi}^{l}_{L}\gamma^{\bar{\mu}}D^{(4)}_{\bar{\mu}}\psi^{l}_{L}+
D^{(4)}_{\bar{\mu}}\bar{\psi}_{lR}\gamma^{\bar{\mu}}\psi_{lR}
-\bar{\psi}_{lR}\gamma^{\bar{\mu}}D^{(4)}_{\bar{\mu}}\psi_{lR}
+D^{(4)}_{\bar{\mu}}\bar{\psi}_{\nu_{l}R}\gamma^{\bar{\mu}}\psi_{\nu_{l}R}
-\bar{\psi}_{\nu_{l}R}\gamma^{\bar{\mu}}D^{(4)}_{\bar{\mu}}\psi_{\nu_{l}R})+\\\\+
\hbar c
kg'a_{L\bar{0}}\bar{\psi}_{L}\gamma^{\mu}B_{\mu}\psi_{L}+\hbar kcg
\sum_{i=1}^{3}a_{L\bar{i}}\bar{\psi}_{L}\gamma^{\mu}W^{i}_{\mu}\psi_{L}+\hbar
kcg'
a_{R\bar{0}}\bar{\psi}_{lR}\gamma^{\mu}B_{\mu}\psi_{lR}\bigg)+\frac{c^{4}}{16\pi
G}R_{N}+\\\\+\sum_{g=1}^{3}\bigg(\frac{i\hbar
c}{2}(D^{(4)}_{\bar{\mu}}\bar{\psi}^{g}_{L}\gamma^{\bar{\mu}}\psi^{g}_{L}
-\bar{\psi}_{L}^{g}\gamma^{\bar{\mu}}D^{(4)}_{\bar{\mu}}\psi^{g}_{L}+
D^{(4)}_{\bar{\mu}}\bar{u}_{gR}\gamma^{\bar{\mu}}u_{gR}
-\bar{u}_{gR}\gamma^{\bar{\mu}}D^{(4)}_{\bar{\mu}}u_{gR}+\\\\+D^{(4)}_{\bar{\mu}}\bar{d}_{gR}\gamma^{\bar{\mu}}d_{gR}
-\bar{d}_{gR}\gamma^{\bar{\mu}}D^{(4)}_{\bar{\mu}}d_{gR}\bigg)+
\hbar kcg'a_{L\bar{0}}\bar{\psi}_{gL}\gamma^{\mu}
B_{\mu}\psi_{gL}+ \end{eqnarray*}
\begin{equation}\label{aztot}
+\hbar kcg
\sum_{i=1}^{3}a_{gL\bar{i}}\bar{\psi}^{g}_{L}\gamma^{\mu}W^{i}_{\mu}\psi^{g}_{L}+\hbar
kcg' a_{uR\bar{0}}\bar{u}_{gR}\gamma^{\mu}B_{\mu}u_{gR}+\hbar kcg'
a_{dR\bar{0}}\bar{d}_{gR}\gamma^{\mu}B_{\mu}d_{gR}\bigg)\bigg]
\end{equation}
with
\begin{equation}
\psi^{l}_{L}=\left(\begin{array}{c}\psi_{\nu_{l}L} \\ \psi_{lL}
\end{array}\right)\qquad\psi^{g}_{L}=\left(\begin{array}{c}u_{gL} \\
d_{gL}\end{array}\right)
\end{equation}
\begin{eqnarray*}
B_{\mu\nu}=\partial_{\nu}B_{\mu}-\partial_{\mu}B_{\nu}\\
W^{i}_{\mu\nu}=\partial_{\nu}W^{i}_{\mu}-\partial_{\mu}W^{i}_{\nu}+kgC^{i}_{jk}W^{j}_{\mu}W^{k}_{\nu}.
\end{eqnarray*}
Imposing that the action (\ref{aztot}) coincides with the
electro-weak standard model one before symmetry breaking, we get
\begin{eqnarray}
\frac{c^{4}}{16\pi G}k^{2}g^{2}\alpha^{2}=1 \qquad \hbar ckg=g\\\\
\frac{c^{4}}{16\pi G}k^{2}g'^{2}\alpha'^{2}=1 \qquad \hbar ckg'=g'
\end{eqnarray}
and thus extra-dimensions lengths are determined by the coupling
constants g and g', according with Weinberg \cite{We83},
\begin{equation}
\alpha^{2}=16\pi G\bigg(\frac{\hbar}{gc}\bigg)^{2}\qquad
\alpha'^{2}=16\pi G\bigg(\frac{\hbar}{g'c}\bigg)^{2};
\end{equation}
by the relations with electric charge and Weinberg angle, we,
finally, have an estimate of the compattification
\begin{equation}
\label{stima}\alpha=0.18\times 10^{-31}cm\qquad
\alpha'=0.33\times10^{-31}cm.
\end{equation}
\\
Therefore, we achieve the geometrization of electro-weak standard
model in a seven-dimensional space-time manifold and we also
obtain an estimate of extra-dimensional lengths which agrees with
the hypothesi that they actually are not observable.

\subsection{Spontaneous symmetry breaking}
At this point we need to reproduce symmetry breaking mechanism; so
we introduce in the theory an adjunctive two components scalar
field $\Phi$ subjected to Higgs potential
\begin{equation}
\label{lhig}\Lambda_{\Phi}=\frac{1}{2}\eta^{(A)(B)}\partial_{(A)}\Phi^{\dag}\partial_{(B)}\Phi
-\mu^{2}\Phi^{\dag}\Phi-\lambda(\Phi^{\dag}\Phi)^{2},
\end{equation}
and with the following dependence by extra-coordinates
\begin{equation}
\label{Phi}\Phi=\frac{1}{\sqrt{V}}e^{-iT_{\bar{i}}\lambda^{\bar{i}}_{\Phi\bar{Q}}\Theta^{\bar{Q}}}\left(\begin{array}{c}e^{-i\lambda^{\bar{0}}_{\Phi1\bar{R}}\Theta^{\bar{R}}}\phi_{1}
\\ e^{-i\lambda_{\Phi2\bar{R}}^{\bar{0}}\Theta^{\bar{R}}}\phi_{2}
\end{array}\right)
\end{equation}
where $\lambda$ coefficients are defined by (\ref{deflam})
with
\begin{equation}
a_{\Phi1\bar{0}}=-\frac{1}{2} \qquad a_{\Phi2\bar{0}}=\frac{1}{2}
\qquad a_{\Phi\bar{i}}=1.
\end{equation}
If we carry out the dimensional reduction of the action we obtain
from (\ref{lhig}), imposing that the functions $\Theta^{\bar{P}}$ satisfy
\begin{equation}
\int d^{3}y\sqrt{-\gamma}\frac{1}{V^{K}}\xi^{m}_{\bar{M}}
\partial_{m}\Theta^{\bar{Q}}\xi^{n}_{\bar{N}}
\partial_{n}\Theta^{\bar{R}}=a_{\Phi1\bar{M}}(\lambda^{-1}_{\Phi1})^{\bar{Q}}_{\bar{M}}
a_{\Phi1\bar{N}}(\lambda^{-1}_{\Phi1})^{\bar{R}}_{\bar{N}}=
a_{\Phi2\bar{M}}(\lambda^{-1}_{\Phi2})^{\bar{Q}}_{\bar{M}}a_{\Phi2\bar{N}}(\lambda^{-1}_{\Phi2})^{\bar{R}}_{\bar{N}}
\end{equation}
\begin{eqnarray*}
\int
e^{m}_{(m)}\partial_{m}\Theta^{\bar{Q}}e^{n}_{(n)}\partial_{n}\Theta^{\bar{P}}\sqrt{-\gamma}d^{3}y=
\frac{1}{\alpha^{\bar{Q}}\alpha^{\bar{P}}}\int
e'^{m}_{(m)}\partial_{m}\Theta^{\bar{Q}}e'^{n}_{(n)}\partial_{n}\Theta^{\bar{P}}\sqrt{-\gamma}d^{3}y=
\frac{1}{\alpha^{\bar{Q}}\alpha^{\bar{P}}}K^{\bar{Q}\bar{P}}_{(m)(n)}
\end{eqnarray*}
with $K^{\bar{Q}\bar{R}}_{\bar{M}\bar{N}}$ constant,
\begin{eqnarray*}
S_{\Phi}=\frac{1}{c}\int d^{4}x \sqrt{-g}
\bigg[\frac{1}{2}g^{\mu\nu}(D_{\mu}\phi)^{\dag}D_{\nu}\phi-
\mu^{2}\phi^{\dag}\phi-\lambda(\phi^{\dag}\phi)^{2}+\frac{1}{2}
G\phi^{\dag}\phi\bigg]
\end{eqnarray*}
where
\begin{equation}
\phi=\left(\begin{array}{c}\phi_{1} \\ \phi_{2}
\end{array}\right)
\end{equation}
and
\begin{equation}
G=\bigg(\sum_{\bar{P}=0}^{3}\frac{1}{(\alpha^{\bar{P}})^{2}}\lambda^{\bar{P}}_{\Phi\bar{Q}}\lambda^{\bar{P}}_{\Phi\bar{S}}+\frac{1}{\alpha^{\bar{0}}\alpha^{\bar{3}}}(\lambda^{\bar{3}}_{\Phi\bar{Q}}\lambda^{\bar{0}}_{\Phi\bar{S}}+\lambda^{\bar{0}}_{\Phi\bar{Q}}\lambda^{\bar{3}}_{\Phi\bar{S}}\bigg)\eta^{\bar{m}\bar{n}}
K^{\bar{R}\bar{S}}_{\bar{m}\bar{n}}.
\end{equation}
Thus, we have after dimensional reduction a four-dimensional field
with the same action as Higgs field;
but, from (\ref{Phi}), the transformation laws for $\phi_{1}$ and
${\phi_{2}}$ under translations along $S^{1}$ are
\begin{eqnarray}\label{trlw1}
\phi_{1}'=\bigg(I-i\frac{1}{2}\delta\omega^{\bar{0}}I\bigg)\phi_{1}\\
\label{trlw2}\phi_{2}'=\bigg(I+i\frac{1}{2}\delta\omega^{\bar{0}}I\bigg)\phi_{2}.
\end{eqnarray}
so they transform in different ways under U(1) and they are
\emph{two hypercharge singlet, not a doublet as in standard model}.
Thus our model predicts that Higgs field has two components with different hypercharge.\\
In this way, we can reproduce invariant mass terms for fermions
by
\begin{equation}
\Lambda_{l\phi}=g_{l}(\bar{\Psi}_{lL}\Phi\Psi_{lR}+\bar{\Psi}_{lR}\Phi^{\dag}\Psi_{lL})+
g_{\nu_{l}}(\bar{\Psi}_{lL}\widetilde{\Phi}\Psi_{lR}+\bar{\Psi}_{lR}\widetilde{\Phi}^{\dag}\Psi_{lL})
\end{equation}
where
\begin{equation}
\widetilde{\Phi}=-i[(\Phi)^{\dag}T_{\bar{2}}]^{T}
\end{equation}
and
\begin{equation}
\bar{\Psi}_{lL}\Phi\Psi_{lR}=\sum_{r=1}^{4}\bar{\Psi}_{lLr}\Phi_{1}\Psi_{lRr}
+\sum_{r=5}^{8}\bar{\Psi}_{lLr}\Phi_{2}\Psi_{lRr}.
\end{equation}
\\
 At this point, spontaneous symmetry breaking is achieved
imposing the following aspectation value on vacuum for $\Phi$
\begin{equation}
\Phi=\left(\begin{array}{c} 0 \\
\frac{1}{\sqrt{2}}(1+\sigma(x^{\mu}))
\end{array}\right)
\end{equation}
which leads to just electric charge conservation.
\\\\
There are, however, in this model problems in reproducing Higgs'
mass: in fact dimensional reduction produces an adjunctive mass
term
\begin{equation}
G\cong\frac{1}{(\alpha^{\bar{P}})^{2}},
\end{equation}
which using (\ref{stima}) is $\approx 10^{35} GeV$ so much greater
than mass estimates ($m_{H}\lesssim 200 GeV$) and, thus, it
imposes an extremely accurate fine-tuning on the parameter
$\mu^{2}$.\\

\subsection{$\alpha$ as Klein-Gordon fields}

Let consider the part of action (\ref{aztot}) that refers to
$\alpha$
\begin{equation}\label{alaz}
S_{\alpha}=\frac{c^{3}}{16\pi G}\int
d^{4}x\sqrt{-g}\bigg(-2g^{\mu\nu}\sum_{\bar{n}=1}^{3}\frac{\nabla_{\mu}\partial_{\nu}\alpha^{\bar{n}}}{\alpha^{\bar{n}}}-g^{\mu\nu}\sum_{\bar{n}\neq
\bar{m}=1}^{3}\frac{\partial_{\mu}\alpha^{\bar{m}}}{\alpha^{\bar{m}}}\frac{\partial_{\nu}\alpha^{\bar{n}}}{\alpha^{\bar{n}}}\bigg);
\end{equation}
where we introduce $\widetilde{A}$
\begin{equation}
\frac{\partial_{\mu}\alpha^{m}}{\alpha^{m}}=\partial_{\mu}\widetilde{A}^{m}\Rightarrow
\widetilde{A}^{m}=\ln\alpha^{m};
\end{equation}
and, by the diagonalization of the second term of (\ref{alaz}),
we get the transformation
\begin{equation}
\left\{\begin{array}{c}A^{1}=\frac{1}{\sqrt{3}}(\widetilde{A}^{1}+\widetilde{A}^{2}+\widetilde{A}^{3})\\\\
A^{2}=\frac{1}{\sqrt{6}}(-\widetilde{A}^{1}+2\widetilde{A}^{2}-\widetilde{A}^{3})\\\\
A^{3}=\frac{1}{\sqrt{2}}(\widetilde{A}^{1}-\widetilde{A}^{3})
\end{array}\right.
\end{equation}
which lead to
\begin{equation}
g^{\mu\nu}\sum_{\bar{n}\neq
\bar{m}=1}^{3}\frac{\partial_{\mu}\alpha^{\bar{m}}}{\alpha^{\bar{m}}}\frac{\partial_{\nu}\alpha^{\bar{n}}}{\alpha^{\bar{n}}}=
g^{\mu\nu}(2\partial_{\mu}A^{1}\partial_{\nu}A^{1}-\partial_{\mu}A^{2}\partial_{\nu}A^{2}-\partial_{\mu}A^{3}\partial_{\nu}A^{3})
\end{equation}
and
\begin{equation}
2g^{\mu\nu}\sum_{\bar{n}=1}^{3}\frac{\nabla_{\mu}\partial_{\nu}\alpha^{\bar{n}}}{\alpha^{\bar{n}}}=g^{\mu\nu}(\sqrt{3}\partial_{\mu}\partial_{\nu}A^{1}+
2\partial_{\mu}A^{1}\partial_{\nu}A^{1}+2\partial_{\mu}A^{2}\partial_{\nu}A^{2}+2\partial_{\mu}A^{3}\partial_{\nu}A^{3}).
\end{equation}
We immediately see that the fields $A^{2}$ and $A^{3}$ are
Klein-Gordon fields, while for $A^{1}$ we define a conformal
transformation on the four-dimensional metric and a transformation
on $A^{1}$ itself
\begin{equation}
g_{\mu\nu}=B^{2}(A^{1})\bar{g}_{\mu\nu}\qquad A^{1}\Rightarrow
D(A^{1})
\end{equation}
where B and D are class $C^{2}$ not zero functions.\\
Lagrangian density in terms of the new metric is
\begin{eqnarray*}
\Lambda=-\frac{c^{4}}{16\pi
G_{(4+K)}}\sqrt{-g}B^{2}V^{K}\bigg[\bar{R}-\frac{1}{4}
\eta_{\bar{m}\bar{n}}\alpha^{\bar{n}}\alpha^{\bar{m}}\bar{F}^{\bar{m}}_{\mu\nu}\bar{F}^{\bar{n}\mu\nu}-\bar{g}^{\mu\nu}\sum_{\bar{n}=2}
^{3}\partial_{\mu}A^{m}\partial_{\nu}A^{m}-\\\\
-2\bar{g}^{\mu\nu}\bar{\nabla}_{\mu}\partial_{\nu}A^{1}\bigg(3\frac{B'}{B}-
2D'\bigg)-\bar{g}^{\mu\nu}\partial_{\mu}A^{1}\partial_{\nu}A^{1}\bigg(4D''-6\frac{B''}{B}+8\frac{B'}{B}D'+5(D')^{2}\bigg)+R_{N}\bigg]
\end{eqnarray*}
and using the degrees of freedom given by the functions B and D we
impose
\begin{equation}
\left\{\begin{array}{c}3\frac{B'}{B}- 2D'=0\\\\
4D''-6\frac{B''}{B}+8\frac{B'}{B}D'+5(D')^{2}=\frac{69}{4}d^{2}
\end{array}\right.
\end{equation}
and thus
\begin{equation}
\left\{\begin{array}{c}D=d A^{1}+g\\
B=fe^{dA^{1}}\end{array}\right.
\end{equation}
\\
with d, f e g arbitrary constants. For the space $S^{1}\otimes
S^{2}$ we have
\begin{equation}
V^{K}\varpropto \alpha^{1}\alpha^{2}\alpha^{3}=e^{\sqrt{3}A^{1}}
\end{equation}
so we can take f and d so that
\begin{equation}
B^{2}V^{K}=1
\end{equation}
and with the transformation
\begin{equation}
A^{1}\Rightarrow \frac{\sqrt{3}}{2}\sqrt{69}A^{1}
\end{equation}
we obtain for the lagrangian density
\begin{equation}
\Lambda=-\frac{c^{4}}{16\pi
G_{(4+K)}}\sqrt{-g}\bigg[\bar{R}-\frac{1}{4}
\eta_{\bar{m}\bar{n}}\alpha^{\bar{n}}\alpha^{\bar{m}}\bar{F}^{\bar{m}}_{\mu\nu}\bar{F}^{\bar{n}\mu\nu}-\bar{g}^{\mu\nu}\sum_{\bar{n}=1}
^{3}\partial_{\mu}A^{n}\partial_{\nu}A^{m}+R_{N}(A^{n})\bigg];
\end{equation}
therefore we can interpret the scalar fields contained in the
extra-dimensional metric as Klein-Gordon fields interacting each
other.\\
This scalar fields may be used to derive spontaneous
symmetry breaking mechanisms, that so would be related to geometric
proprieties of space-time and would not require the introduction of an
adjunctive scalar field and the fine-tuning on its potential parameters.\\
We also observe that the number of $\alpha$ fields is equal to the
number of degrees of freedom necessary to attribute transversal
components and so masses to three gauge bosons even if we do not
know a spontaneous symmetry breaking mechanism able to achieve
this aim, in fact in standard electro-weak model Higgs fields has
got four independent components.

\section{Conclusions}

In this work we presented the generalization of Kaluza-Klein theories to general compact
homogeneous extra-dimensional spaces in order
to geometrize a generic gauge group theory. \\
Beside the geometrization of the bosonic component, well-know in literature,
the main results of our work have been \begin{itemize}
\item the identification of gauge transformations with extra-dimensional
translations and the interpretation of gauge charges as
extra-dimensional momentum components, so that the proprieties of
interaction of fields descend from their dependence
from extra-coordinates; \item the geometrization of gauge connections for spinors,
that lead to introduce matter by free spinors.\end{itemize}
Therefore, by starting from Einstein-Hilbert plus Dirac action the dimensional splitting gives a four-dimensional theory that describes gravity, gauge bosons, spinorial fields and their interactions; however to eliminate some adjunctive free currents terms we had to assume not-standard spinorial connections.\\
In comparing the model with known theories, we eliminated extra-coordinates by an integration, whose explanation is  related to not-observability of extra-dimensions.\\
Then, as an application to a gauge group U(1) we described the original pentadimensional
Kaluza-Klein theory in presence of free spinorial matter fields with
a phase dependence on the fifth coordinate; in this way we got the issue to geometrize QED.\\
We also showed how, by a conformal transformation on the four dimensional metric, it is possible to interpret
the not well-known scalar field of the original theory as a Klein-Fock one and then we reproduced the theory with spinors after the conformal transformation.\\
We also presented the application to the electro-weak model
and thus to a non-abelian gauge group: we considered a space-time
manifold $V^{4}\otimes S^{1}\otimes S^{2}$ and we reproduced all
standard model particle by two spinors for every leptonic
family and quark generation, with suitable dependence from
extra-coordinates in order to have the correct interaction
proprieties and to overall the chirality problem. We also obtained
from coupling constants an estimate of extra-dimensions lengths.
Finally to achieve spontaneous symmetry breaking we showed how the
dimensional splitting of a scalar field with two opposite
hypercharge components could lead to the Higgs field plus
invariant mass terms for fermions, even if we need an extremely
accurate fine-tuning on the parameter in the potential that
suggested us to look for alternative ways. We retain that one of
this could be related to the scalar fields in the
extra-dimensional metric that can be interpreted as
Klein-Gordon fields.\\\\
However, in this kind of theories there are problems that concern
the rule of extra-dimensional spinorial connections and the
explanation of the compactification and of the breaking of general
invariance, for which a both solution we retain could be find
by 
spontaneous compactification mechanisms.\\
Moreover the formulation of a quantistic theory of gravity, because of compattification, become, if possible, more important when we introduce extra-dimensions; in particular models based on non commutative geometries for the adjunctive space \cite{Co85}
\cite{Ma96} seems to tend toward this issue and they also could reproduce in a geometrical way spontaneous symmetry breaking
mechanism \cite{OK95}.\\
Finally Kaluza-Klein theories' prospectives are applications to more complex gauge group in order to geometrize
strong interaction or Grand Unification scheme.\\

\end{document}